\documentclass[12pt]{article}

\usepackage{graphicx}
\usepackage{epsfig,cite}
\usepackage{amsmath}
\usepackage{amssymb}
\usepackage{color}
\usepackage{latexsym,url}
\usepackage{pdflscape}

\def\jhep #1 #2 #3 {{JHEP} {\bf #1} (#2) #3}
\def\plb #1 #2 #3 {{Phys.~Lett.} {\bf B#1} (#2) #3}
\def\npb #1 #2 #3 {{Nucl.~Phys.} {\bf B#1} (#2) #3}
\def\epjc #1#2 #3 {{Eur.~Phys.~J.} {\bf C#1} (#2) #3}
\def\epjd #1#2 #3 {{Eur.~Phys.~J.} {\bf D#1} (#2) #3}
\def\zpc #1 #2 #3 {{Z.~Phys.} {\bf C#1} (#2) #3}
\def\jpg #1 #2 #3 {{J.~Phys.} {\bf G#1} (#2) #3}
\def\prd #1 #2 #3 {{Phys.~Rev.} {\bf D#1} (#2) #3}
\def\prep #1 #2 #3 {{Phys.~Rep.} {\bf #1} (#2) #3}
\def\prl #1 #2 #3 {{Phys.~Rev.~Lett.} {\bf #1} (#2) #3}
\def\mpl #1 #2 #3 {{Mod.~Phys.~Lett.} {\bf #1} (#2) #3}
\def\rmp #1 #2 #3 {{Rev. Mod. Phys.} {\bf #1} (#2) #3}
\def\cpc #1 #2 #3 {{Comp. Phys. Commun.} {\bf #1} (#2) #3}
\def\sjnp #1 #2 #3 {{Sov. J. Nucl. Phys.} {\bf #1} (#2) #3}
\def\xx #1 #2 #3 {{\bf #1}, (#2) #3}
\def\hepph #1 {{\tt hep-ph/#1}}

\newcommand{\be}{\begin{equation}}

\newcommand{\ee}{\end{equation}}
\newcommand{\bea}{\begin{eqnarray}}
\newcommand{\eea}{\end{eqnarray}}
 
\newcommand{\smartpap}{p\hskip-7pt\hbox{$^{^{(\!-\!)}}$}}

\newcommand{\smallz}{{\scriptscriptstyle Z}} %
\newcommand{\mz}{m_\smallz}
\newcommand{\smallw}{{\scriptscriptstyle W}}
\newcommand{\mw}{m_\smallw} 
\newcommand{\mwc}{m_{\smallw 0}} 
\newcommand{\mwi}{m_{\smallw, i}} 
\newcommand{\mwj}{m_{\smallw, j}} 
\newcommand{\mwjbar}{m_{\smallw, \bar j}} 
\newcommand{\gw}{\Gamma_{\smallw}} 
\newcommand{\gz}{\Gamma_{\smallz}}

\newcommand{\sdw}{\sin^2\theta_{\smallw}}

\begin{document}

\newenvironment{appendletterA}
 {
  \typeout{ Starting Appendix \thesection }
  \setcounter{section}{0}
  \setcounter{equation}{0}
  \renewcommand{\theequation}{A\arabic{equation}}
 }{
  \typeout{Appendix done}
 }

\begin{titlepage}
\nopagebreak

\renewcommand{\thefootnote}{\fnsymbol{footnote}}
\vskip 1.5cm
\begin{center}
\boldmath
{\Large\bf PDF uncertainties on the W boson mass measurement}\\[9pt]
{\Large\bf from the lepton transverse momentum distribution}\\[9pt]
\unboldmath
\vskip 1.cm
{\large G. Bozzi\footnote{Email: giuseppe.bozzi@mi.infn.it}}\\
{\it Dipartimento di Fisica, Universit\`a degli Studi di Milano,\\
Via Celoria 16, 
I-20133 Milano, Italy} \\[5mm]

{\large L. Citelli\footnote{Email: luca.citelli@studenti.unimi.it}}\\
{\it Dipartimento di Fisica, Universit\`a degli Studi di Milano,\\
Via Celoria 16, 
I-20133 Milano, Italy} \\[5mm]

{\large A.~Vicini\footnote{Email: alessandro.vicini@mi.infn.it}}\\
{\it Dipartimento di Fisica, Universit\`a degli Studi di Milano and
INFN, Sezione di Milano,\\
Via Celoria 16, 
I-20133 Milano, Italy} \\[5mm]

\end{center}

\begin{abstract}
We study the charged current Drell-Yan process and we evaluate the proton parton densities uncertainties on the lepton transverse momentum distribution
and their impact on the determination of the $W$ boson mass.
We consider the global PDF sets {\tt CT10, MSTW2008CPdeut, NNPDF2.3, NNPDF3.0, MMHT2014},
and apply the PDF4LHC recipe to combine the individual results, obtaining 
an uncertainty on $\mw$ that ranges between $\pm 18$ and $\pm 24$ MeV, depending on the final state, collider energy and kind.
We discuss the dependence of the uncertainty on the acceptance cuts
and the role of the individual parton densities in the final result.
We remark that some PDF sets predict an uncertainty on $\mw$ of ${\cal O}(10{\rm MeV})$; this encouraging result is spoiled, in the combined analysis of the different sets, by an important spread of the central values predicted by each group.
\end{abstract}

\vfill
\end{titlepage}    

\setcounter{footnote}{0}
\tableofcontents
\clearpage
%%%%%%%%%%%%%%%%%%%%%%%%%%%%%%%%%%%%%%%%%%%%%%%%%%%%%%%%%%%%%%%%%%%%%%%%%%%%%%
\section{Introduction}

The very accurate measurement of the $W$ boson mass performed at the Tevatron experiments CDF ($\mw=80.387\pm 0.019$ GeV) \cite{Aaltonen:2013vwa} and D0 ($\mw=80.375\pm 0.023$ GeV) \cite{D0:2013jba}, with a world average now equal to $\mw=80.385\pm 0.015$ GeV \cite{{Beringer:1900zz}}, offers the possibility of a high-precision test of the gauge sector of the Standard Model (SM).
There are prospects of a further reduction of the total experimental uncertainty, with a final error of ${\cal O}(10)$ MeV 
for the combination of LHC, Tevatron and LEP results \cite{Buge:2006dv,Besson:2008zs}.

The current best prediction in the SM is $\mw=80.357\pm 0.009 \pm 0.003$ GeV \cite{Degrassi:2014sxa}
and has been computed including the full 2-loop corrections \cite{Awramik:2003rn}, augmented by higher-order QCD corrections
\cite{fourloopQCD}
 and by resumming reducible contributions.
%
%and partial three-loop (${\cal O}(\alpha\alpha_{s}^{2})$, ${\cal O}(\alpha_{t}^{2}\alpha_{s})$, ${\cal O}(\alpha_{t}^{3})$) and four-loop QCD corrections ${\cal O}(\alpha_{t}\alpha_{s}^{3})$, where $\alpha_{t}=\alpha m_{t}^{2}$. 
The uncertainty on this evaluation is mostly due to parametric uncertainties of the inputs of the calculation, the top mass value, the hadronic contribution to the running of the electromagnetic coupling, and to theoretical uncertainties.

The simultaneous indirect determination of the top quark mass and the $W$ mass, together with the direct determination of the Higgs boson mass, provides an important consistency check for the Standard Model \cite{Baak:2014ora};
in turn the comparison of an accurate experimental $M_{W}$ measurement with the predictions of different models might provide an indirect signal of physics beyond the SM.

The $W$ boson mass is extracted by means of a template fit technique applied to different observables of the charged-current (CC) Drell-Yan (DY) process, namely the lepton and neutrino transverse-momenta ($p_{T}^{l}$, $p_{T}^{\nu}$) and the lepton pair transverse mass $m_{T}$ (see, for instance, \cite{Aaltonen:2013vwa,D0:2013jba}).
The differential distributions are computed with Monte Carlo simulation codes for different values of $\mw$ and are subsequently compared with the corresponding data: the value which maximizes the agreement is chosen as preferred value for $\mw$.

The present CDF and D0 results are affected by a systematic error
obtained as the combination of several elements, both of experimental and theoretical origin (see Tables IX and X in ref. \cite{Aaltonen:2013vwa} for CDF and Table VI in ref. \cite{D0:2013jba} for D0).

Among the experimental items, the most problematic ones are 
the determination of the lepton energy scale (${\cal O}(7-17)$ MeV) and of the recoil scale and resolution (${\cal O}(5-8)$ MeV),
where the recoil $\vec u_T$ is defined as the sum of the momenta of all the measured charged tracks, with the exception of the ones associated to the lepton(s).
The mismeasurement of the recoil affects the determination of the $W$ transverse momentum and the application of the selection cuts on this variable; in turn it affects the determination of the leptons transverse momenta, with an impact on the final $\mw$ value.
The largest contribution of theoretical systematic error is due to the parametrization of the proton parton density functions (PDFs), 
which will be the main subject of the present paper,
while another theoretical item present in these tables is the size of the missing QED effects not included in the available simulation tools,
estimated to be of ${\cal O}(4)$ MeV.

The recoil modeling is an important element of the analysis of charged-current DY events, for it enters in the determination of the neutrino and of the charged lepton transverse momenta.
The model is validated on neutral-current DY data of the lepton-pair transverse momentum distribution, thanks to the fact that,
in this latter case, the full information about the kinematics in the transverse plane can be reconstructed.
The propagation of this calibration of the recoil model to the $\mw$ measurement is estimated to be of ${\cal O}(3)$ MeV.
The description of the recoil was optimized at the Tevatron experiments in the
region of small lepton-pair transverse momenta, so that the DY events used for the $\mw$ determination are eventually selected imposing a cut $\vec u_T < 15$ GeV. In the following we will discuss the impact of this selection criterium, which will implemented in our analysis as a cut on the transverse momentum of the lepton-neutrino pair.

The theoretical contribution to the final systematic error 
enters in the analysis of the data via the templates.
In fact the latter are computed with codes based on perturbative calculations, which are truncated at a finite order in the expansion parameter. 
The proton PDFs, which are used to describe the partonic content of the proton, are affected by the error of the data from which they are extracted; this error propagates to the templates, inducing an additional systematic uncertainty of the fitting tool.
The interplay between the $W$ and $Z$ transverse momentum distributions and the proton PDFs has been discussed in \cite{Brady:2011hb,Brandt:2013hoa}, without a discussion on the consequence for the $\mw$ determination.

The aim of the present paper is to provide a quantitative assessment of the error induced by our imperfect knowledge of the proton PDFs, in the preparation of the templates used to fit $\mw$, in the case of the charged-lepton transverse momentum ($p_{T}^{l}$) distribution,
as it is measured in the hadron collider processes $p\smartpap \to W^\pm \to l\nu+X$. 

In \cite{Bozzi:2011ww} an analysis based on the lepton-pair transverse mass shows that PDF uncertainties do not challenge a measurement of $M_{W}$ at 10 MeV accuracy. 
As discussed above, the measurement of the lepton $p_T$ has different, 
and to a certain extent complementary, systematic uncertainties, 
compared to the lepton-pair transverse mass.
On the other hand, getting theoretical predictions for the $p_{T}^{l}$ distribution can be quite challenging due to the high sensitivity of this exclusive observable to the details of the description of the radiation emitted. 
In this respect, also the PDFs parametrization has a direct impact on the shape of the distribution and, in turn, on $\mw$.

\subsection{The lepton transverse momentum distribution and $\mw$ }
\begin{figure}[!h]
\begin{center}
\includegraphics[width=80mm,angle=0]{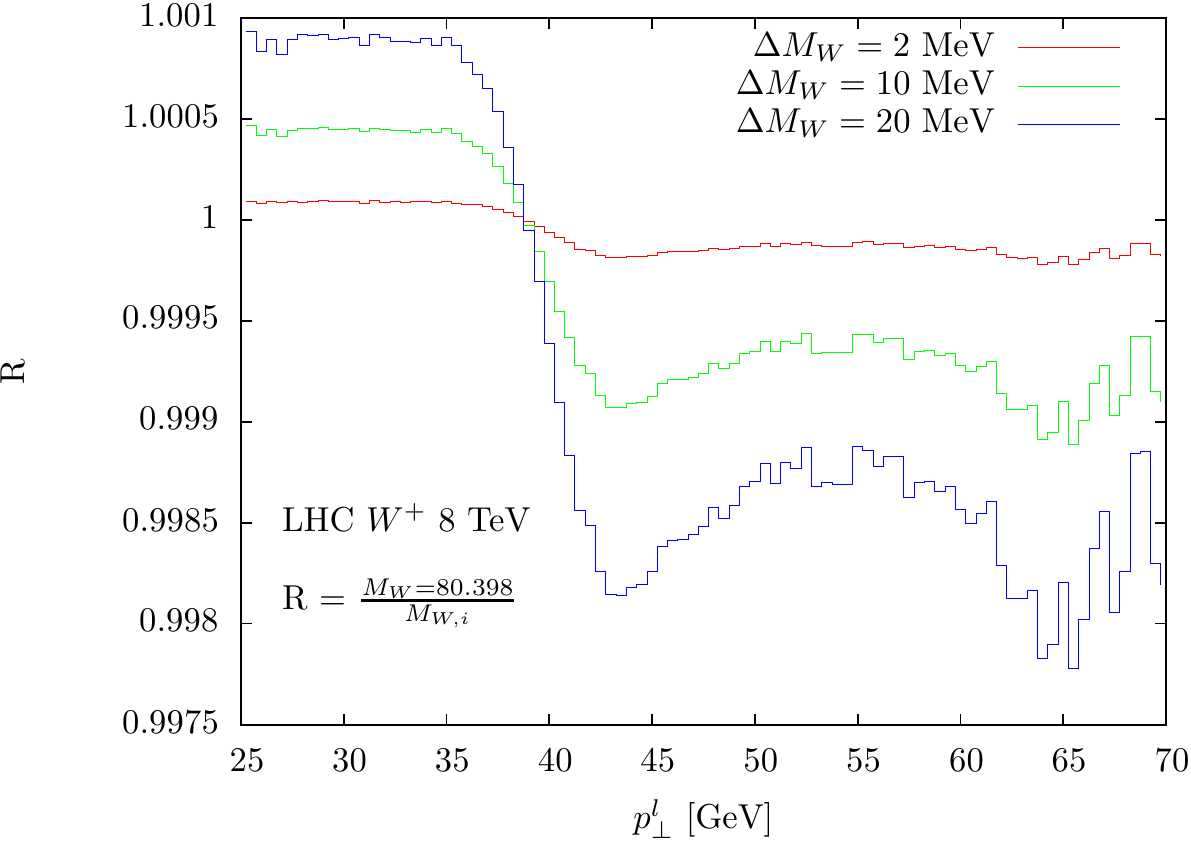}
\caption{ 
Ratio of lepton transverse momentum distributions
which have been generated with different $W$ boson masses.
\label{fig:sensitivity}
}
\end{center}
\end{figure}

The sensitivity of the lepton transverse momentum distribution to the precise value of the $W$ boson mass
is due to its jacobian peak which has its maximum for $p_\perp^l\sim\mw/2$.
A change of $\mw$ in the simulation codes by $2,10,20$ MeV, with respect to a fixed reference value,
yields a distortion of the distribution in the per mill range, as illustrated in Figure \ref{fig:sensitivity}.
Any effect (perturbative QCD corrections, PDF uncertainties, etc.) that induces a change of the shape of this distribution of similar size
represents a source of systematic theoretical uncertainty on the $\mw$ determination; in particular a measurement at the 10 MeV level
requires a control of the shape of the templates at the 1 per mill level or better.

There are two mechanisms that yield a distribution of the lepton transverse momentum in the DY processes: the decay of the gauge boson and its recoil against QCD (and in smaller amount QED) radiation.
The initial state radiation collinear divergences make any fixed-order prediction for this quantity unreliable,
because of the important contributions in the region of small gauge boson transverse momenta,
which have to be resummed to all orders.
It is thus necessary to use a code that implements the resummation to all orders of multiple gluon emissions, either analytically or in a numerical approach via Parton Shower (PS), to obtain a physically sensible prediction.
In this study we use the {\tt POWHEG} Monte Carlo event generator \cite{Alioli:2008gx}, matched with the PYTHIA \cite{Sjostrand:2006za} QCD PS.
The accuracy of this code on the inclusive DY cross section is NLO-QCD, while from the point of view of the enhancement due to the logarithms of the transverse momentum of the lepton pair, the lepton transverse momentum distribution has LL accuracy.
We consistently choose NLO-QCD PDF distributions.
The {\tt POWHEG} 
event generator is currently used by the ATLAS  and CMS (see e.g. respectively refs. \cite{Aad:2014qja} and \cite{CMS:2014jea})
collaborations to study the DY processes and provides a good description of the data. We thus consider it as a valid starting point to study the propagation of the PDF uncertainty in the preparation of the templates eventually used to fit the data.

A more accurate description of the charged-lepton transverse momentum distribution can be obtained by the use of codes that include higher-order QCD corrections, like {\tt ResBos} \cite{Balazs:1997xd,Berge:2004nt} 
or {\tt DYRes} \cite{Bozzi:2008bb},
or EW and QED multiple photon effects, like {\tt POWHEG} \cite{Bernaciak:2012hj,Barze:2012tt,Barze':2013yca}.
As it is well known, final state QED radiation plays a crucial role in the precise determination of MW \cite{CarloniCalame:2003ux}. 
However, since in this study our main focus is on the assessment of the impact on the $\mw$ determination of the PDF uncertainty,
we choose a fast code that yields a basic realistic description of the shape of this distribution. The difference with the predictions that one could obtain adopting one of the other above listed codes belongs to the class of mixed PDF $\times$ higher-order effects and it can be estimated as a perturbative correction to the results of the present study; 
the use of a code that includes final state QED effects modifies the basic 
shape of the templates, with a shift of the central $\mw$ value of ${\cal O}(200)$ MeV \cite{homero}; 
our {\it ansatz}, in the absence of an explicit check, is that
this modification of the shape yields also a rescaling of all the PDF uncertainties at most of ${\cal O}(10\%)$ of the results discussed in this paper; this estimate follows from the comparison between the QED shift and the size of the region of the $p_\perp^l$ distribution sensitive to an $\mw$ variation, which is of at least of one, but more realistically of a few $W$ decay widths.
We would thus obtain a change in our results, for a given PDF set,
by ${\cal O}$(2 MeV),
which is comparable to the statistical accuracy that we can claim in the template fit and to the error, of experimental origin, on the PDF uncertainty itself \cite{Demartin:2010er}.

The plan of the paper is the following: 
in Section \ref{sec:tools} we recall the definition of some basic theoretical tools which will be used in the study; 
in Section \ref{sec:numres} we present our numerical results, discussing the PDF uncertainty on the lepton transverse-momentum distribution and on the $\mw$ determination; we also consider the dependence of the $\mw$ PDF uncertainty on the acceptance cuts and comment on possible future developments;
in Section \ref{sec:conclusions} we draw our conclusions.
%%%%%%%%%%%%%%%%%%%%%%%%%%%%%%%%%%%%%%%%%%%%%%%%%%%%%%%%%%%%%%%%%%%%%%%%
\section{Theoretical tools}
\label{sec:tools}
In this section, we briefly outline the strategy adopted to estimate the PDF uncertainty in the determination of $m_{W}$ at hadron colliders: we refer the interested reader to \cite{Bozzi:2011ww} for more details.

\subsection{Template fit}
\label{sec:template}
In this paper we discuss the uncertainty on $\mw$ of different PDF sets.
In order to make a quantitative evaluation, we follow some basic steps:
\begin{enumerate}
\item
we generate the lepton transverse momentum ${\cal O}$ with different PDF replicas, keeping the $W$ mass fixed at a given value 
$\mwc$, and we treat each distribution as a set of {\it pseudodata};
\item
we compute the {\it templates}, i.e. the distributions that are used to fit the pseudodata,
with one specific choice for the PDF set ({\tt NNPDF2.3}, replica 0) and we let $\mw$ assume all the values in the interval
$[80.312,80.470]$ in steps of 2 MeV;
\item
we compare a given pseudodata distribution, obtained with a given PDF replica labelled by $i$, 
with all the templates labelled by $j$; in each comparison we compute an indicator
\be
\chi_{i,j}^2=
\frac{1}{N_{\rm bins}}
\sum_{k=1}^{N_{\rm bins}}
\frac{\left(\mathcal{O}_k^{j, template}-\mathcal{O}_k^{
      i,data}\right)^2}{(\sigma_k^{i,data})^2+(\sigma_k^{j,template})^2-2{\rm Cov}(data,template)}
\label{eq:chi2j}
\ee
where ${\cal O}_k$ and $\sigma_k$ are respectively the value of the distribution and its associated error in the bin $k$ and Cov is the covariance between the two distributions;
\item
the template $\bar j$ that yields the minimum value of $\chi_{i,j}^2$ is the one that best describes the pseudodata and its associated $\mwjbar$ value is thus the preferred value associated to the replica $i$; 
the difference $\Delta \mwi=\mwjbar -\mwc$, is the shift induced by the PDF replica $i$ chosen for that set of pseudodata;
in other words it is the difference between the results that we would obtain when fitting the real data if we prepared the templates with the replica $i$ instead of the replica 0 of {\tt NNPDF2.3}. 

\end{enumerate}

\subsection{PDF uncertainties}
\label{sec:pdfunc}
The proton PDF sets considered in this study are
{\tt MSTW2008CPdeut}~\cite{Martin:2009iq}, 
{\tt CT10}~\cite{Gao:2013xoa}, 
{\tt NNPDF2.3}~\cite{Ball:2012cx},
{\tt NNPDF3.0}~\cite{Ball:2014uwa} and
{\tt MMHT2014}~\cite{Harland-Lang:2014zoa}.
They are called global sets because they include all the available relevant hard scattering data. 
Each collaboration provides a prescription to estimate the PDF uncertainties
\footnote{We refer to the original publications for more details.}: 
in particular we recall the formula for the symmetric error in the Hessian approach ({\tt CT10,  MSTW2008CPdeut, MMHT2014})
for a generic observable $X$
\be
\Delta X = \frac12 \sqrt{\sum_{i=1}^{N_{eigenvectors}}\left[X^+_i-X^-_i \right]^2  }
\label{eq:hessian}
\ee
where the sum runs over the $N_{eigenvectors}$ eigenvectors in parameter space, with the associated pairs of replicas (+ and -).
Instead with {\tt NNPDF} the average and the standard deviation over the ensemble $\{q\}$ of $N_{rep}$ PDF replicas
provide the estimate of the best value and of the error on the observable ${\cal F}$:
\be
\langle\mathcal{F}
[\{q\}]\rangle=\frac{1}{N_{rep}}\sum_{k=1}^{N_{rep}}\mathcal{F}[\{q^{(k)}\}]\,\, ,
\label{eq:mcmean}
\ee
\be
\sigma_{\mathcal{F}}=\left(\frac{1}{N_{rep}-1}
\sum_{k=1}^{N_{rep}}\left(\mathcal{F}[\{q^{(k)}\}]-
\langle\mathcal{F}[\{q\}]\rangle\right)^2\right)^{1/2}\,\, . 
\label{eq:mcstd}
\ee
The results obtained with these three PDF sets can be combined according to the current PDF4LHC recommendation~\cite{Botje:2011sn},
to find a conservative estimate of the PDF uncertainty.

In this paper we apply this procedure to two observables, namely the lepton transverse momentum distribution and the $W$ mass determined with the template fit procedure.

\subsection{Correlation functions}
\label{sec:correlation}
A useful quantity to evaluate the role of the different parton densities in the hadronic cross section is
the correlation function $\rho$ 
between the parton-parton luminosities
and the charged-lepton distribution at a given value of the transverse momentum.
The parton-parton luminosity is defined as
${\cal P}_{ij}(x,\tau)=f_i(x,\mu_F^2) f_j(\frac{\tau}{x},\mu_F^2)$ 
where $f_i(x,\mu_F^2)$ is the density describing a parton $i$ at a scale $\mu_F$
and $\tau=\frac{M^2}{S}$ with $M$ the final state invariant mass and $S$ the hadronic Mandelstam invariant.
The correlation $\rho$ is defined as
\be
\rho(x,\tau) = 
\frac{ \langle {\cal P}_{ij}(x,\tau) \frac{d\sigma}{dp_\perp^l}   \rangle
-\langle {\cal P}_{ij}(x,\tau)  \rangle
\langle \frac{d\sigma}{dp_\perp^l} \rangle   }
{\sigma^{PDF}_{{\cal P}_{ij} } \sigma^{PDF}_{d\sigma/dp_\perp^l}}\, ,
\label{eq:correlation}
\ee
where the angle brackets indicate average with respect to the different PDF replicas.
%In Figure \ref{fig:pdfunccuts2}
%we show the correlation function $\rho$ for $\tau=10^{-4}$ at the LHC, with fixed $p_\perp^l=40.5$ GeV.

\clearpage
%%%%%%%%%%%%%%%%%%%%%%%%%%%%%%%%%%%%%%%%%%%%%%%%%%%%%%%%%%%%%%%%%%%%%%%%
\section{Numerical results}
\label{sec:numres}

%%%%%%%%%%%%%%%%%%%%%%%%%%%%%%%%%%%%%%%%%%%%%%%%%%%%%%%%%%%%%%%%%%%%%%%%%
\subsection{Input parameters and setup}
We simulate the processes $p\smartpap \to W^+ \to \mu^+\nu_\mu+X$  and $p\smartpap \to W^- \to \mu^-\bar\nu_\mu+X$ 
in proton-antiproton collisions with $\sqrt{S}= 1.96$ TeV
and 
in proton-proton collisions with $\sqrt{S}= 8, 13, 33, 100$ TeV energies.
In the absence of QED effects, not considered here, our results will be identical to those obtained with electrons instead of muons. 
We consider the PDF sets 
{\tt MSTW2008CPdeut} \cite{Martin:2009iq},
{\tt CT10} \cite{Gao:2013xoa}, 
{\tt NNPDF2.3} \cite{Ball:2012cx},
{\tt NNPDF3.0} \cite{Ball:2014uwa},
{\tt MMHT2014} \cite{Harland-Lang:2014zoa},
and use the corresponding values of
$\alpha_s(\mz)$.
We use the following values for the input parameters in the Monte Carlo codes:
\begin{center}
\begin{tabular}{lll}
$G_{\mu} = 1.16637~10^{-5}$ GeV$^{-2}$ & 
$\mw = 80.398$~GeV&
$\mz=91.1876$~GeV\\
$\sdw = 1 - \mw^2/\mz^2$&
$\gw = 2.141$~GeV & 
$\gz = 2.4952$~GeV \\
$V_{cd}=0.222$ &
$V_{cs}= 0.975$ &
$V_{cb}=0$ \\
$V_{ud}=0.975$ &
$V_{us}=0.222$ &
$V_{ub}=0$ \\
$V_{td}=0$ &
$V_{ts}=0$ &
$V_{tb}=1$ \\
\end{tabular}
\end{center}
The charm quark  in the partonic cross section is treated as a massless particle, while the bottom quark does not contribute because of the vanishing top density in the proton.
As for the kinematic cuts, we used those summarized in Table~\ref{tab:selection}, similar to those used in the corresponding experimental analysis: the main difference between the Tevatron and LHC is the wider acceptance for the rapidity of the leptons in the latter case. 
The $p_{T}^{l}$ distribution has been studied in the interval $29$ GeV$\leq p_\perp^l\leq 49$ GeV, with a bin size of 0.5 GeV. All the following analyses are performed with bare leptons both in the pseudodata and in the templates. 
\begin{table}[!h]
\begin{center}
\begin{tabular}{|c|c|}
\hline
Tevatron & LHC \\
\hline
$p_{\perp}^{\mu} \geq$~25 GeV        & $p_{\perp}^{\mu} \geq$~25 GeV \\
$\rlap{\slash}{\! E_T} \geq$~25 GeV &  $\rlap{\slash}{\! E_T} \geq$~25
GeV \\ 
$|\eta_\mu|< 1.0$                   & $|\eta_\mu|< 2.5$ \\
$p_\perp^W<15$ GeV & $p_\perp^W<15$ GeV \\
\hline
\end{tabular}
\caption{\small Selection criteria for DY $W\to l\nu$ events
for the Tevatron and the LHC.}
\label{tab:selection}
\end{center}
\end{table}

The Monte Carlo simulation requires a specific, technical comment.
The effects under study are deformations of the shape of the lepton transverse momentum distribution at the per mill level, either due to a variation of the $\mw$ value or to a different PDF replica choice.
This distribution receives contributions from a large fraction of the available final-state phase space, making very difficult an accurate dedicated sampling. As a consequence, Monte Carlo statistical fluctuations at the per mill level are present also with hundreds of milions of simulated unweighted events.
The solution to this problem is found using a reweighting technique, based on the remark that both the dependence on the PDFs and the dependence on $\mw$ factorize from the rest of the fixed-order partonic cross section.
Only one simulation, one sequence of events is used to generate all the templates and all the pseudodata: the weight $w_0$ associated to each event is corrected by an appropriate reweighting factor to account for different replica or, separately, $\mw$ value choices,
\bea
w_0&\to& w_j = w_0 \frac{(\hat s-\mwc^2)^2 + \gw^2 \mwc^2}{(\hat s-\mwj^2)^2 + \gw^2 \mwj^2} ~~~~~~~~~~~~\text{template~~}j\\
w_0&\to& w_i = w_0 \frac{f_i(x_1) g_i(x_2)}{f_0^{NNPDF}(x_1) g_0^{NNPDF}(x_2)} \nonumber~~~~~~~~~~~~~\text{replica}~~i
\eea
where $f,g$ are two generic parton densities.
In the {\tt POWHEG} formulation this rescaling spoils the exact NLO accuracy of the final result, by terms generated by the {\tt POWHEG} Sudakov form factor. The size of the latter could not be distinguished from Monte Carlo statistical fluctuations, when we compared two distributions, one obtained with an exact simulation and the other introducing the new PDF replica via the above rescaling. The main results of this study should thus not be affected.
An update of the {\tt POWHEG} generator is in progress \cite{nv}
to restore the NLO accuracy after reweighting.

Since the events used  are exactly the same, the statistical fluctuations of the different distributions (templates and pseudodata) are highly correlated
(correlation is about 0.987, almost constant over the bins, with 80M of events) and cancel to a large extent when we compute the difference of two cross sections in a bin.
%% The Monte Carlo statistical uncertainty that affects the fitting procedure
%% can be estimated by reading the width of the $\chi^2$ parabola of Equation \ref{eq:chi2j}, applying the $\Delta\chi^2=1$ rule. 
%% This procedure has just a qualitative value, because a rigorous statistical interpretation is possible only when estimating the actual value of a paramater from a sample of data generated with the same model used to prepare the templates.
A statistical error of $\pm 2$ MeV can be obtained with the simulation of 1620 millions of events.
We checked that, with increasing statistics, the result of each individual fit is stable, because the uniform reduction of the statistical error in the different $p_\perp^l$ bins.
We stress that, rather than an absolute prediction of the $\mw$ value, we are interested in the quantitative assessment: $i)$ of the relative difference between various PDF sets and $ii)$ of the PDF uncertainty within one set, which is defined as the spread with respect to a central value. In both cases we have to provide a solid estimate of differences bin by bin and the reweighting procedure allows to efficiently remove the Monte Carlo fluctuation effects, leaving only the physically relevant shifts.

%%%%%%%%%%%%%%%%%%%%%%%%%%%%%%%%%%%%%%%%%%%%%%%%%%%%%%%%%%%%%%%%%%%%%%%%%%%
\subsection{PDF uncertainty of the distribution }
We study the percentage PDF uncertainty
on the lepton transverse momentum distribution and also on the associated
normalized distribution defined as
\be
\frac{d\bar\sigma}{dp_\perp^l} = 
\frac{1}{\left( \int_{p_\perp^{min}}^{p_\perp^{max}} dp_\perp^l ~\frac{d\sigma}{dp_\perp^l} \right) } \cdot
\frac{d\sigma}{dp_\perp^l}  
\label{eq:normdistr}
\ee
As it is well known from the Tevatron experiments \cite{Aaltonen:2013vwa,D0:2013jba},
the uncertainty is in general reduced in the normalized observable.
In fact, each PDF replica in a set contributes in a different way to the shape and to the overall normalization of the physical distributions;
by considering the normalized distributions of Equation \ref{eq:normdistr} we are sensitive mostly to the shape change;
the latter is the most relevant item in the determination of $\mw$, because we associate the precise position and shape of the jacobian peak to the value of the gauge boson mass.
\begin{figure}[!h]
\includegraphics[width=85mm,angle=0]{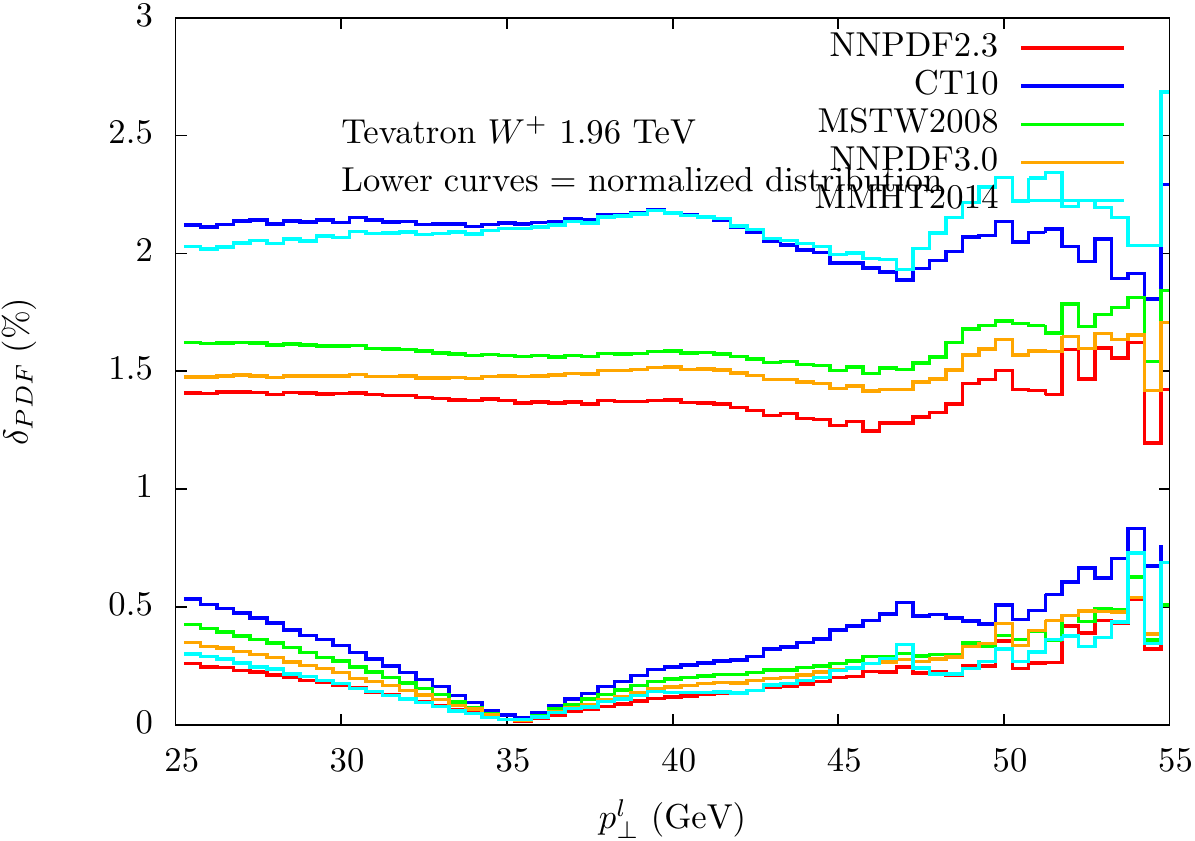}
\hfill\\
\includegraphics[width=85mm,angle=0]{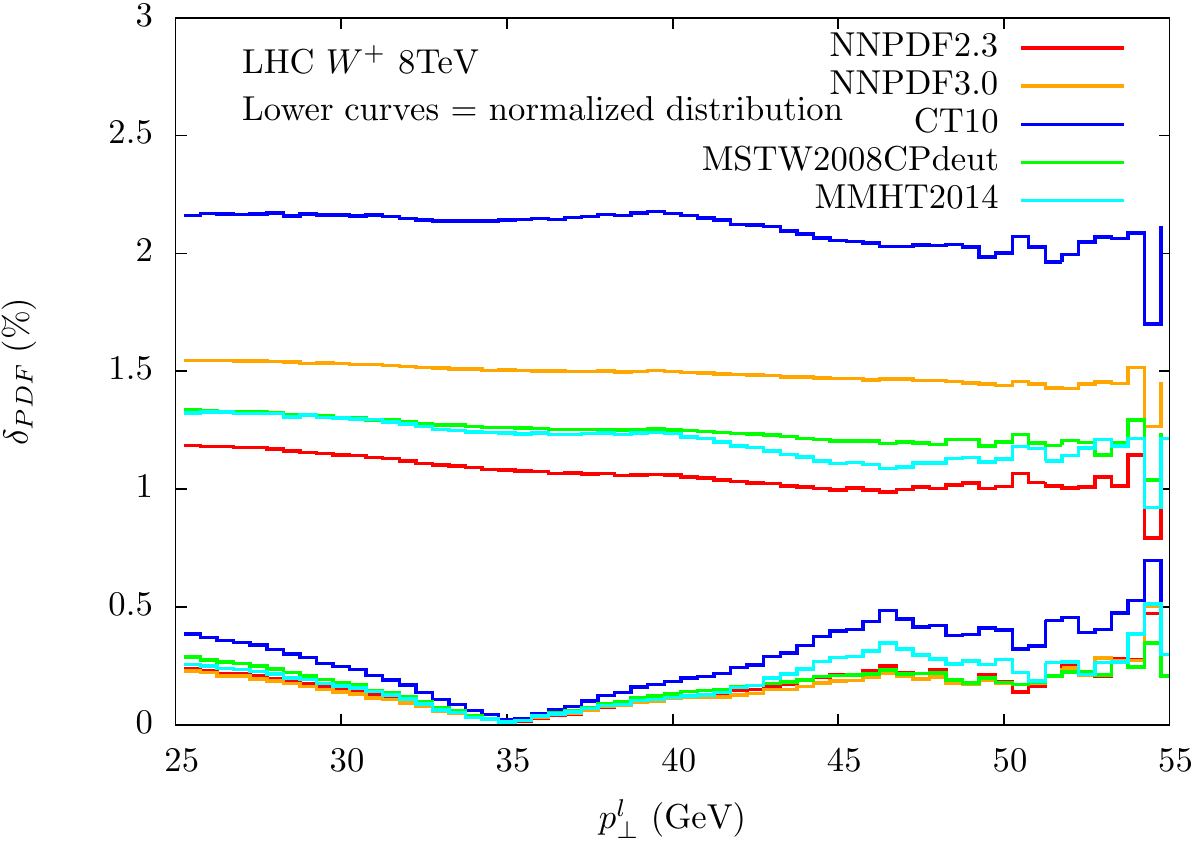}
\includegraphics[width=85mm,angle=0]{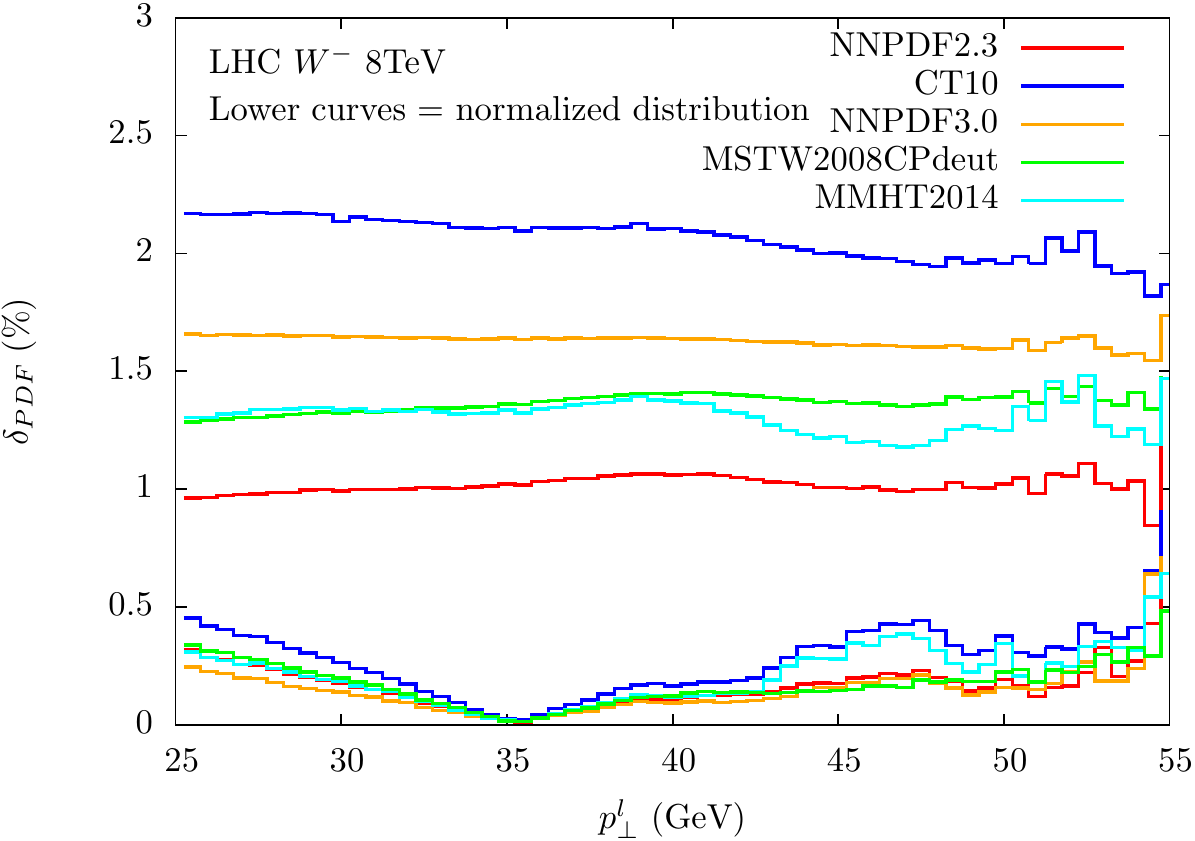}\\
\includegraphics[width=85mm,angle=0]{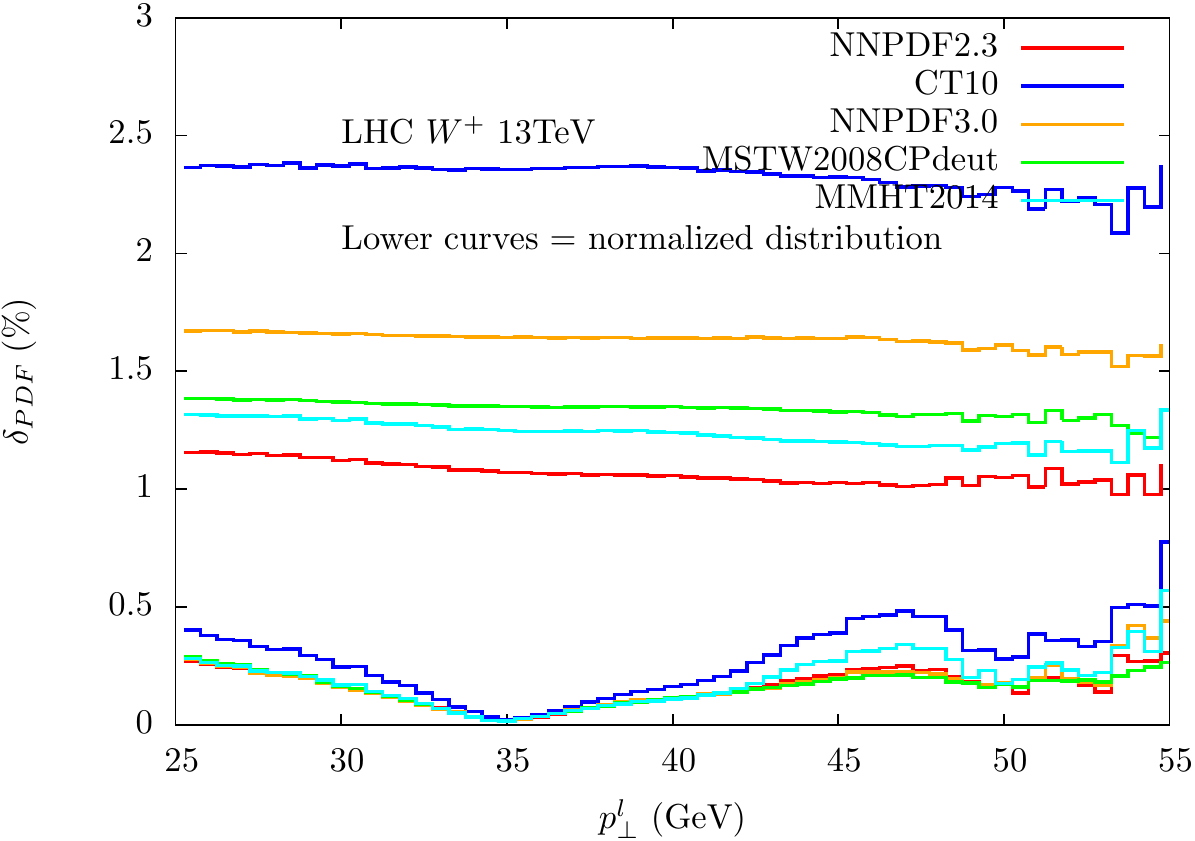}
\includegraphics[width=85mm,angle=0]{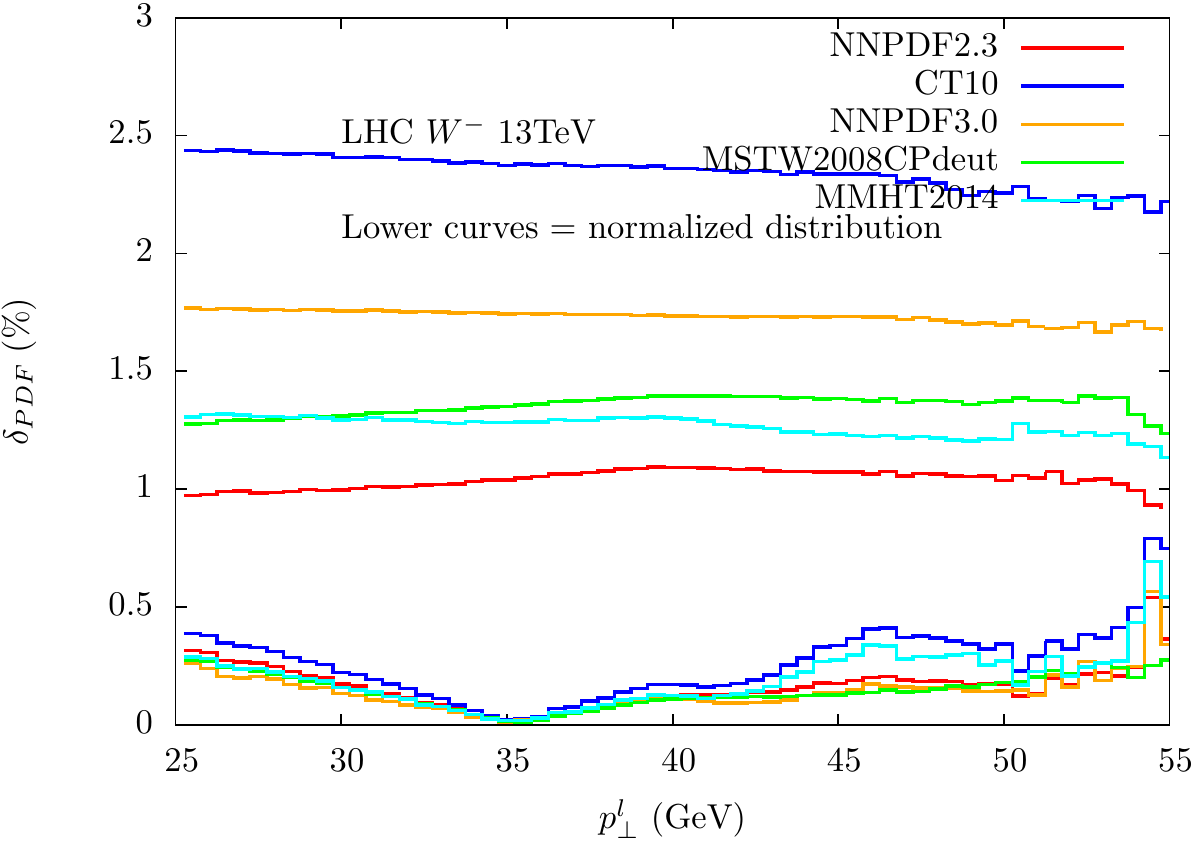}
\caption{ 
Percentage size of the PDF uncertainty on the lepton transverse momentum distribution, computed with different PDF sets.
%with the {\tt NNPDF2.3} (solid red), {\tt CT10} (dashed blue) and {\tt MSTW2008CPdeut} (dotted green) sets,
In addition to the basic acceptance criteria of Table \ref{tab:selection}, a cut  $p_\perp^W< 15$ GeV on the lepton pair has been applied.
The lower lines refer to the normalized distributions of Equation \ref{eq:normdistr}, the upper lines to the standard ones.
Results for $W^+$ (left) and $W^-$ (right) production
at the LHC 8 TeV (middle plots) and 13 TeV (lower plots); results for the Tevatron in the upper plot.
\label{fig:pdfunc-sets}
}
\end{figure}
\begin{figure}[!h]
\includegraphics[width=80mm]{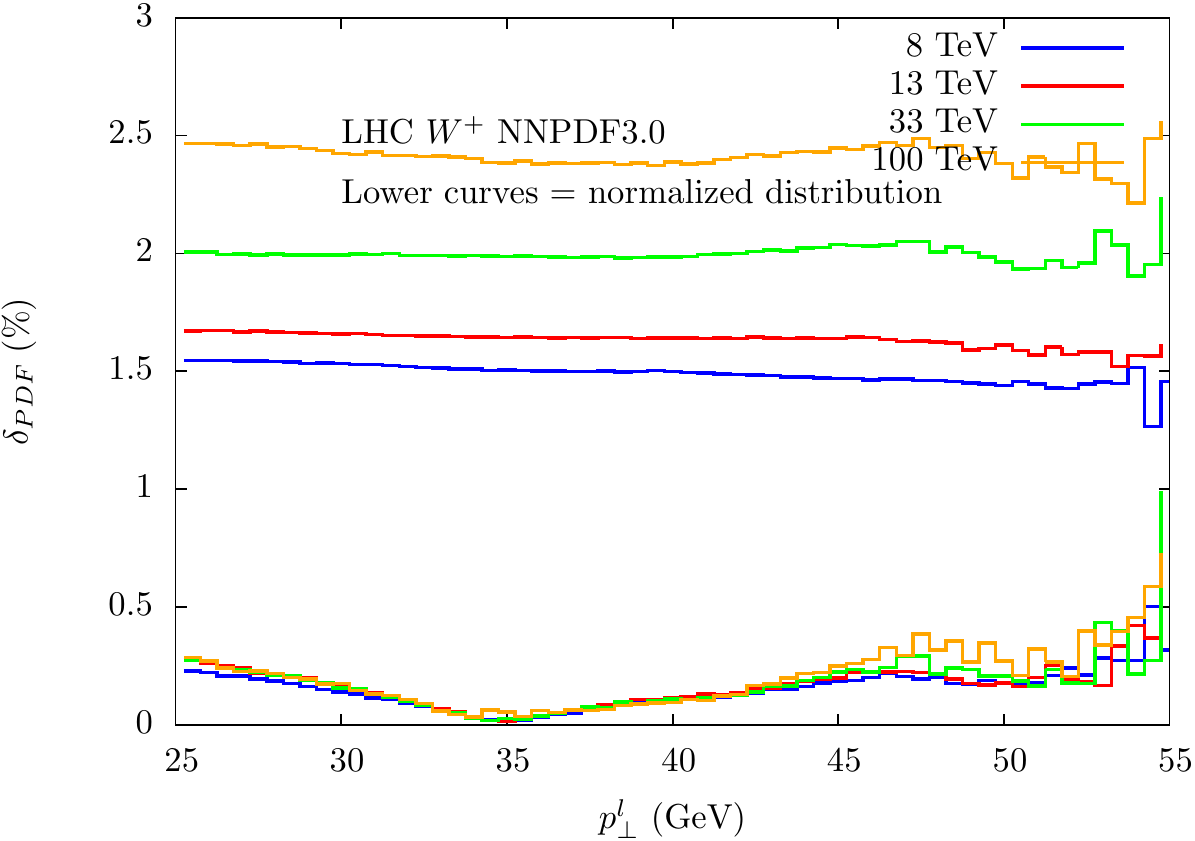}
\includegraphics[width=80mm]{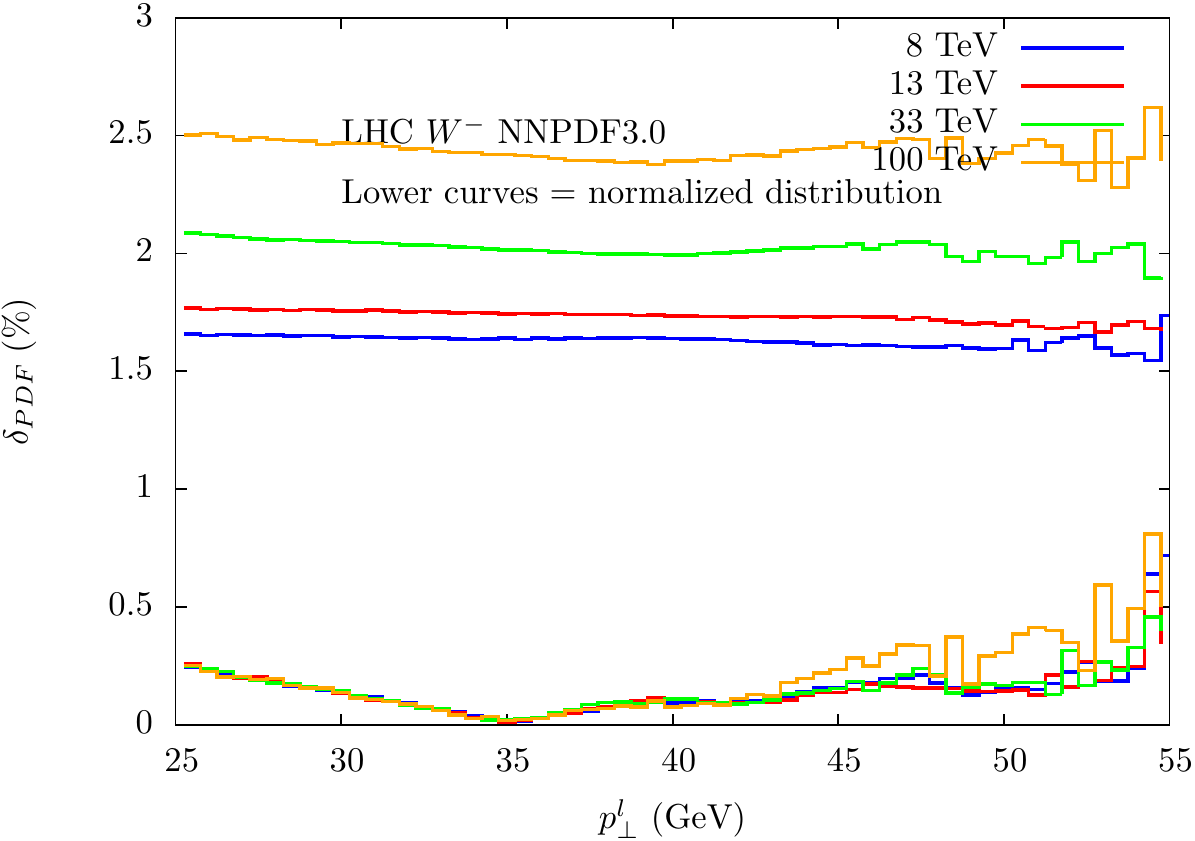}
\caption{ 
Percentage size of the PDF uncertainty on the lepton transverse momentum distribution,
computed with the {\tt NNPDF3.0} set at the LHC at different energies.
Results for $W^+$ (left) and $W^-$ (right) production, in the case of absolute (upper lines) or normalized (lower lines) distributions.
\label{fig:pdfunc-energy}
}
\end{figure}
In Figure \ref{fig:pdfunc-sets} we show the PDF uncertainty of the lepton transverse momentum distribution and of the associated normalized distribution,
computed at the Tevatron, at the LHC 8 and 13 TeV with different PDF sets, both in the cases of $W^+$ and $W^-$ production, in presence of the additional cut $p_\perp^W<15$ GeV with respect to the choices indicated in Table \ref{tab:selection}.
The percentage uncertainty of the normalized distributions is at the few per mill level at the jacobian peak, 
and could mimic the effect of a $\mw$ shift by ${\cal O}(10)$ MeV.
In Figure \ref{fig:pdfunc-energy} we use the PDF set {\tt NNPDF3.0} and study the change of the PDF uncertainty with the collider energy,
in presence of the additional cut $p_\perp^W<15$ GeV.
The uncertainty of the distribution increases, as function of the collider energy, from 1.5 to 2.5\%,
while in the normalized case the uncertainty is almost independent of the energy.

%%%%%%%%%%%%%%%%%%%%%%%%%%%%%%%%%%%%%%%%%%%%%%%%%%%%%%%%%%%%%%%%%%%%%%
\subsection{Impact of the PDF uncertainty on the $\mw$ determination }
\label{sec:mwspread}
\begin{figure}[!h]
\begin{center}
\includegraphics[width=80mm,angle=0]{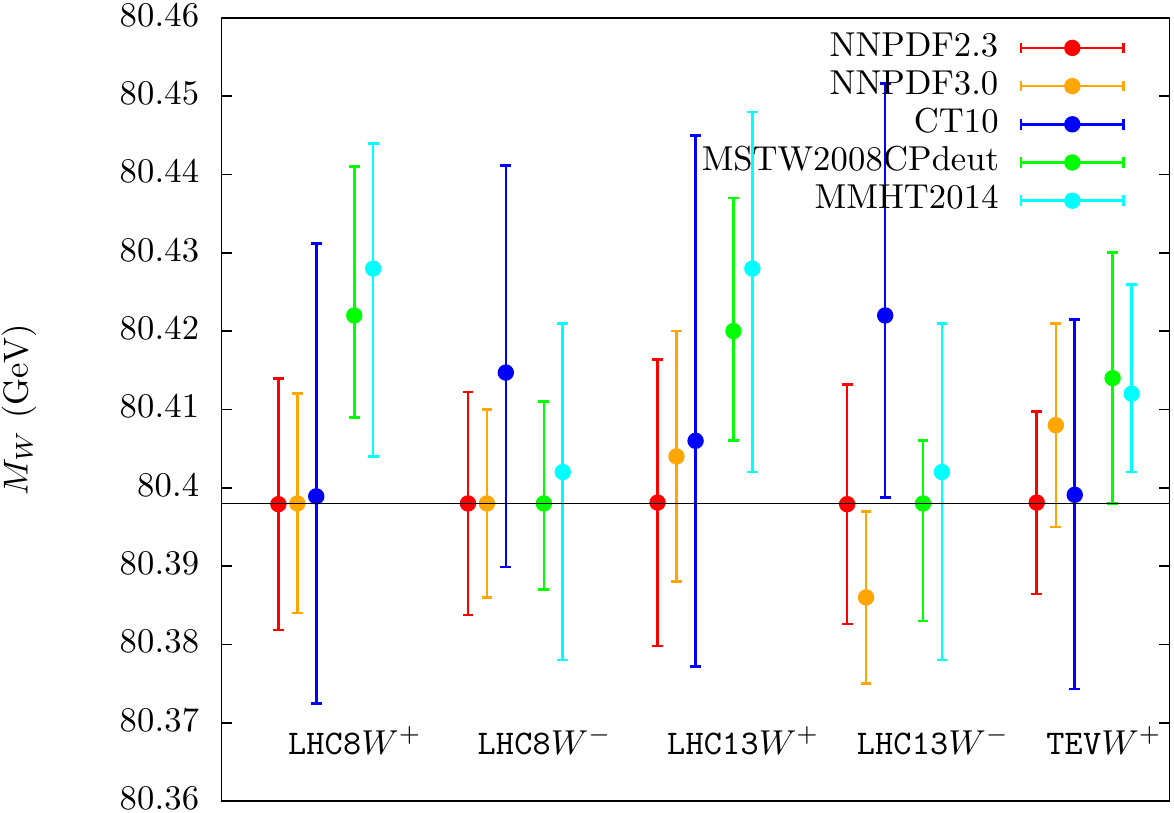}
\includegraphics[width=80mm,angle=0]{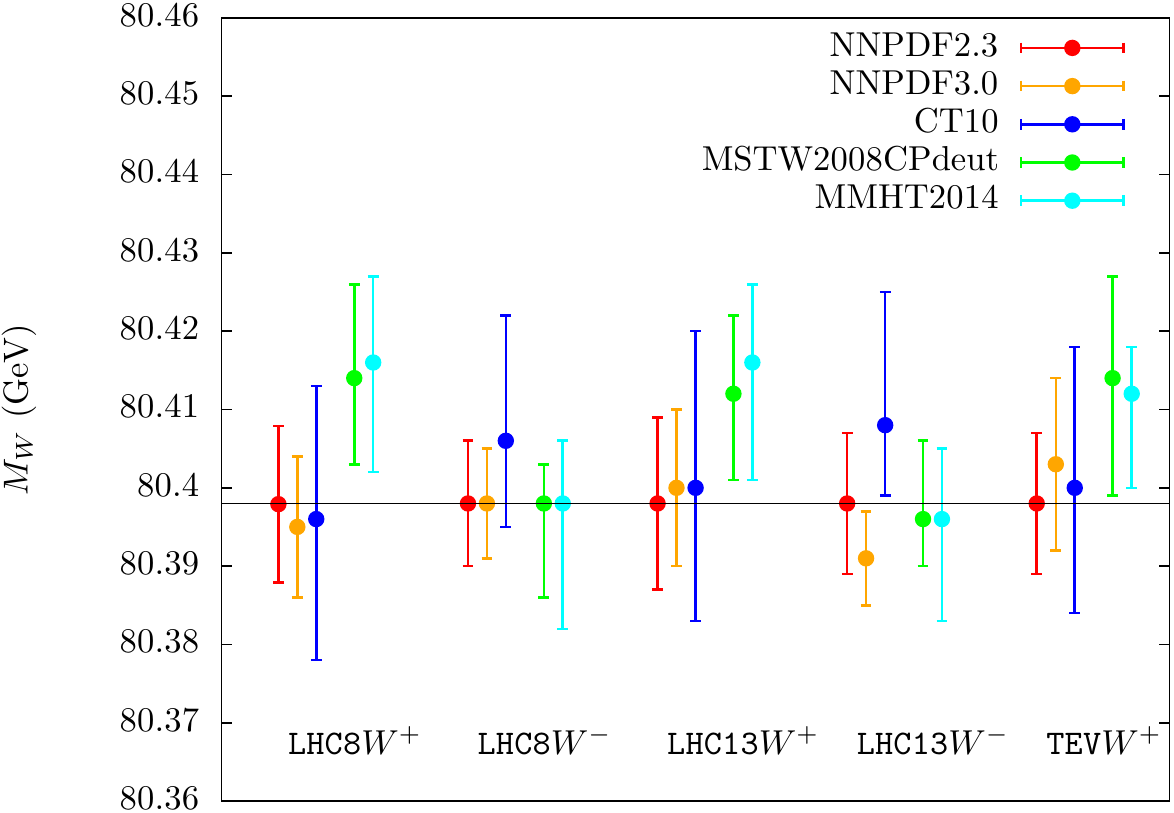}
\caption{
Summary of the PDF uncertainty on $\mw$ computed with different PDF sets, colliders and final states.
The basic acceptance criteria have been used in the left plot, while in the right plot an additional cut $p_\perp^W<15$ GeV has been applied.
\label{fig:summaryunc}
}
\end{center}
\end{figure}

The template fit procedure, described in Section \ref{sec:template},
has been applied to the distributions computed with all the replicas of the different PDF sets under study;
the corresponding preferred $\mw$ values have been combined, 
according to the rules described in Section \ref{sec:pdfunc},
to derive the uncertainty on the $\mw$ extraction due to the PDFs.
The fit interval has been chosen to be $p_\perp^l \in [29,49]$ GeV,
in order to minimize the contribution to the PDF uncertainty from the tails of the distribution above and below the jacobian peak.
The template fit has been applied to our pseudodata generated with a fixed value of the $W$ boson decay width $\gw$.
We have checked that our results are weakly dependent on the choice of this parameter: 
we repeated the fit using for $\gw$ a value modified by $\pm \sigma_\Gamma$, 
where $\sigma_\Gamma=0.042$ GeV is the current experimental error,
and we found that the prediction for the PDF uncertainty on $\mw$ gets modified by 1-2 MeV, depending on the selection cuts.
The results for the Tevatron and for the LHC 8 and 13 TeV 
are presented in Table \ref{tab:mwunc1} 
%and \ref{tab:mwunc2}
and
are also summarized in Figures \ref{fig:summaryunc}.
In the upper half of Table \ref{tab:mwunc1} (and in Figure \ref{fig:summaryunc} left plot) 
no additional cut on $p_\perp^W$ has been imposed on the lepton pair, 
whereas in the lower half of the same Table (and in Figure \ref{fig:summaryunc} right plot) 
a cut $p_\perp^W<15$ GeV has been applied.

The PDF uncertainty reflects the experimental error of the data from which the parton densities are extracted and also the different methodologies used in their determination.
As it can be observed from Figure \ref{fig:summaryunc},
the estimate of the PDF uncertainty on $\mw$ predicted by the different PDF collaborations differs by a factor up to 3 between the different groups.
The uncertainty on $\mw$ extracted from normalized distributions, 
with the basic selection criteria of Table \ref{tab:selection}, ranges 
from 12 to 23 MeV at the Tevatron, 
from 12 to 29 MeV at the LHC 8 TeV and 
from 11 to 34 MeV at the LHC 13 TeV.
Imposing on the lepton pair a cut $p_\perp^W<15$ GeV modifies these results; the ranges of the PDF uncertainties become:
from 11 to 17 MeV at the Tevatron, 
from 7 to 17 MeV at the LHC 8 TeV and 
from 6 to 18 MeV at the LHC 13 TeV.

In addition, the PDF sets under study differ in the parametrization that they adopt to describe the proton structure;
the latter affects the best description of the PDFs and in turn the best prediction of the central $\mw$ value.
The spread $\Delta_{sets}$ of the central values, defined as the difference between the largest and the smallest central values,
is a second component of the final PDF uncertainty on $\mw$. 

A conservative estimate of the uncertainty on $\mw$,
that combines the two elements of uncertainty described above,
can be obtained by computing the envelope of the predictions under study, 
according to the PDF4LHC recipe \cite{Botje:2011sn} 
and by measuring the half-width $\delta_{PDF}$ of the resulting band.
We include, in the evaluation of the envelope, the results of the sets
{\tt CT10}, {\tt MSTW2008CPdeut} and {\tt NNPDF2.3}, because they are based on the same sets of data, making their comparison homogeneous.
These results are presented in Table \ref{tab:spread}.
We observe that the spread $\Delta_{sets}$ represents a large contribution, up to 35\% of the overall uncertainty .
In Table \ref{tab:spread2} we compute the envelope of the results obtained with two more modern PDF sets, namely {\tt NNPDF3.0} and {\tt MMHT2014}, which include recent public data from the LHC.
We observe that the width of the envelope ranges between 16 and 32 MeV, depending on the collider energy and kind and on the final state;
more interesting, the spread of the two central values is below 5 MeV in the
$W^-$ case at the LHC, while it is above 15 MeV in the $W^+$ case and at the Tevatron.

From Table \ref{tab:mwunc1} we can appreciate the impact of the inclusion of the new LHC data, which have been used in the determination of the {\tt NNPDF3.0} set. Beside a few MeV offset for the central values, it is possible to observe a small (few MeV) reduction of the PDF uncertainty, which is roughly 20\% smaller than the one computed with {\tt NNPDF2.3}.
For {\tt MMHT2014} the uncertainties are similar or slightly larger than the ones obtained with {\tt MSTW2008CPdeut}.

The dependence of the PDF uncertainty with the collider energy
is illustrated in Table \ref{tab:mwuncenergy}, using the {\tt NNPDF3.0} PDF set.

\begin{table}[!h]
\begin{center}
\begin{tabular}{|r|c|c|c|c|}
\hline
& \multicolumn{2}{|c|}{no $p_\perp^W$ cut } & \multicolumn{2}{|c|}{$p_\perp^W<15$ GeV  } \\
\hline
  & $\delta_{PDF}$ (MeV) & $\Delta_{sets}$ (MeV)  & $\delta_{PDF}$ (MeV) & $\Delta_{sets}$ (MeV) \\
\hline
Tevatron 1.96 TeV  & 27 & 16 & 21 &  15  \\
\hline
LHC 8 TeV  $W^+$   & 33 & 26 & 24 &  18  \\
\hline
           $W^-$   & 29 & 16 & 18 &  8  \\
\hline
LHC 13 TeV $W^+$   & 34 & 22 & 20 &  14  \\
\hline
           $W^-$   & 34 & 24 & 18 &  12  \\
\hline
\end{tabular}
\caption{
Half-width $\delta_{PDF}$ of the envelope of the PDF uncertainty intervals by {\tt CT10}, {\tt MSTW2008CPdeut} and {\tt NNPDF2.3}.
Corresponding spread $\Delta_{sets}$ of the central predictions.
\label{tab:spread}
}
\end{center}
\end{table}
\begin{table}[!h]
\begin{center}
\begin{tabular}{|r|c|c|c|c|}
\hline
& \multicolumn{2}{|c|}{no $p_\perp^W$ cut } & \multicolumn{2}{|c|}{$p_\perp^W<15$ GeV  } \\
\hline
  & $\delta_{PDF}$ (MeV) & $\Delta_{sets}$ (MeV)  & $\delta_{PDF}$ (MeV) & $\Delta_{sets}$ (MeV) \\
\hline
Tevatron 1.96 TeV  & 16 &  4 & 13 & 9  \\
\hline
LHC 8 TeV  $W^+$   & 32 & 33 & 21 & 21  \\
\hline
           $W^-$   & 22 &  6 & 12 &  0  \\
\hline
LHC 13 TeV $W^+$   & 30 & 24 & 18 & 16  \\
\hline
           $W^-$   & 23 & 16 & 11 &  5  \\
\hline
\end{tabular}
\caption{
Same as in Table \ref{tab:spread},
now considering only the two recent PDF sets 
{\tt NNPDF3.0} and {\tt MMHT2014}.
\label{tab:spread2}
}
\end{center}
\end{table}
\begin{table}[!h]
\begin{center}
\begin{tabular}{|c|c|c|c|c|}
\hline
   \multicolumn{5}{|c|}{normalized distribution, additional cut $p_\perp^W<15$ GeV} \\
\hline
& 8 TeV & 13 TeV & 33 TeV & 100 TeV \\
\hline
$W^+$ & $80.395 \pm 0.009$ & $80.400 \pm 0.010 $ & $80.402 \pm 0.010 $  & $80.404 \pm 0.013 $ \\
\hline
$W^-$ & $80.398 \pm 0.007$  & $80.391 \pm 0.006 $ & $80.385 \pm 0.007 $ & $80.398 \pm 0.011 $ \\
\hline
\end{tabular}
\caption{
Estimate of the central values and of the PDF uncertainty on $\mw$, extracted from the lepton transverse momentum distributions simulated with the {\tt NNPDF3.0} set at different proton-proton collider energies. 
\label{tab:mwuncenergy}
}
\end{center}
\end{table}

\begin{landscape}
\begin{table}[!h]
\begin{center}
\begin{tabular}{|r|c|c|c|c|c|}
\hline
   \multicolumn{6}{|c|}{absolute distributions} \\
\hline
collider/channel & CT10                     &  MSTW2008CPdeut            & NNPDF2.3             & NNPDF3.0         & MMHT2014\\
\hline
Tevatron,  $W^+$ & $80.406 + 0.043 - 0.046 $ &  $80.428 + 0.025 - 0.017$ & $80.400 \pm 0.030$  & $80.427 \pm 0.018$ & $80.430 + 0.022 - 0.022$  \\
\hline
LHC 8 TeV, $W^+$ & $80.394 + 0.040 - 0.029$ &  $80.422 + 0.025 - 0.016$  & $80.398 \pm 0.020$  & $80.406 \pm 0.019$ & $80.428 + 0.027 - 0.022$  \\
\hline
           $W^-$ & $80.444 + 0.055 - 0.062$ &  $80.390 + 0.038 - 0.036 $ & $80.398 \pm 0.030$  & $80.441 \pm 0.027 $ & $80.404 + 0.041 - 0.048$  \\
\hline
LHC 13 TeV,$W^+$ & $80.396 + 0.045 - 0.034$ &  $80.416 + 0.020 - 0.020$  & $80.398 \pm 0.022$  & $80.414 \pm 0.022$ & $80.422 + 0.030 - 0.024$  \\
\hline
           $W^-$ & $80.416 + 0.088 - 0.065$ &  $80.374 + 0.044 - 0.033 $ & $80.398 \pm 0.031$  & $80.426 \pm 0.037 $  & $80.384 + 0.037 - 0.049$  \\
\hline\hline
\multicolumn{6}{|c|}{normalized distributions} \\
\hline
collider/channel & CT10                     & MSTW2008CPdeut             & NNPDF2.3            & NNPDF3.0          & MMHT2014   \\
\hline
Tevatron, $W^+$  & $80.400 + 0.022 - 0.025$ & $80.414 + 0.016 - 0.016$   & $80.398 \pm 0.012$ & $80.408 \pm 0.013$ &  $80.412 + 0.014 - 0.010$ \\
\hline
LHC 8 TeV, $W^+$ & $80.398 + 0.032 - 0.026$ & $80.424 + 0.014 - 0.019$   & $80.398 \pm 0.016$ & $80.395 \pm 0.014$ &  $80.428 + 0.016 - 0.024$ \\
\hline
 $W^-$           & $80.416 + 0.026 - 0.025$ & $80.398 + 0.011 - 0.014$   & $80.398 \pm 0.014$ & $80.396 \pm 0.012$ & $80.402 + 0.019 - 0.024$  \\
\hline
LHC 13 TeV,$W^+$ & $80.406 + 0.039 - 0.029$ & $80.420 + 0.017 - 0.014$   & $80.398 \pm 0.018$ & $80.404 \pm 0.016$ & $80.428 + 0.020 - 0.026$  \\
\hline
 $W^-$           & $80.422 + 0.030 - 0.023$ & $80.398 + 0.008 - 0.015$   & $80.398 \pm 0.015$ & $80.386 \pm 0.011$ & $80.402 + 0.019 - 0.024$ \\
\hline
\hline
   \multicolumn{6}{|c|}{absolute distributions, additional cut $p_\perp^W<15$ GeV} \\
\hline
collider/channel & CT10                     &   MSTW2008CPdeut            & NNPDF2.3             & NNPDF3.0         & MMHT2014 \\
\hline
Tevatron,  $W^+$ & $80.412 + 0.024 - 0.024 $ &   $80.424 + 0.018 - 0.017$ & $80.398 \pm 0.013$ & $80.420 \pm 0.014$ & $80.426 + 0.009 - 0.021$  \\
\hline
LHC 8 TeV, $W^+$ & $80.392 + 0.026 - 0.021$ &   $80.414 + 0.020 - 0.011$  & $80.398 \pm 0.014$ & $80.403 \pm 0.014$ & $80.418 + 0.019 - 0.017$  \\
\hline
           $W^-$ & $80.422 + 0.039 - 0.034$ &   $80.394 + 0.019 - 0.023$  & $80.398 \pm 0.017$ & $80.423 \pm 0.017$ & $80.400 + 0.023 - 0.028$  \\
\hline
LHC 13 TeV,$W^+$ & $80.392 + 0.028 - 0.022$ &   $80.410 + 0.012 - 0.016$  & $80.398 \pm 0.014$ & $80.408 \pm 0.014$ & $80.414 + 0.016 - 0.019$  \\
\hline
           $W^-$ & $80.408 + 0.042 - 0.037$ &   $80.386 + 0.019 - 0.021$  & $80.398 \pm 0.016$ & $80.410 \pm 0.018$ & $80.388 + 0.021 - 0.025$  \\
\hline\hline
\multicolumn{6}{|c|}{normalized distributions, additional cut $p_\perp^W<15$ GeV} \\
\hline
collider/channel & CT10                     &   MSTW2008CPdeut              & NNPDF2.3            & NNPDF3.0         & MMHT2014 \\
\hline
Tevatron, $W^+$  & $80.400 + 0.018 - 0.016$ &   $80.414 + 0.013 - 0.015$    & $80.398 \pm 0.009$ & $80.403 \pm 0.011$ & $80.412 + 0.006 - 0.012$ \\
\hline
LHC 8 TeV, $W^+$ & $80.396 + 0.017 - 0.018$ &   $80.414 + 0.012 - 0.011$    & $80.398 \pm 0.010$ & $80.395 \pm 0.009$ & $80.416 + 0.011 - 0.014$  \\
\hline
 $W^-$           & $80.406 + 0.016 - 0.011$ &   $80.398 + 0.005 - 0.012$    & $80.398 \pm 0.010$ & $80.398 \pm 0.007$ & $80.398 + 0.008 - 0.016$  \\
\hline
LHC 13 TeV,$W^+$ & $80.400 + 0.020 - 0.017$ &   $80.412 + 0.010 - 0.011$    & $80.398 \pm 0.011$ & $80.400 \pm 0.010$ & $80.416 + 0.010 - 0.015$  \\
\hline
 $W^-$           & $80.408 + 0.017 - 0.009$ &   $80.396 + 0.010 - 0.006$    & $80.398 \pm 0.009$ & $80.391 \pm 0.006$ & $80.396 + 0.009 - 0.013$  \\
\hline
\end{tabular}
\caption{
Estimate of the central values and of the PDF uncertainty on $\mw$, extracted from the lepton transverse momentum distributions simulated with different PDF sets and acceptance cuts.
The templates have been generated with {\tt NNPDF2.3} replica 0. The pseudodata for the different PDF sets have been simulated by setting $\mw=80.398$ GeV.
\label{tab:mwunc1}
}
\end{center}
\end{table}

\end{landscape}

\clearpage
%%%%%%%%%%%%%%%%%%%%%%%%%%%%%%%%%%%%%%%%%%%%%%%%%%%%%%%%%%%%%%%%%%%%%%%%
\subsection{PDF uncertainty dependence on the acceptance cuts}
\label{sec:cuts}
The results presented in Section \ref{sec:mwspread} have been obtained imposing on the leptons the basic cuts 
of Table \ref{tab:selection}.
The dependence of the $\mw$ PDF uncertainty on additional cuts on the lepton-pair transverse momentum $p_\perp^W$
or on the charged-lepton pseudorapidity acceptance interval is presented in Table \ref{tab:cuts}.
This study suggests possible optimizations of the event selection, to minimize the PDF uncertainty impact.
It also offers a link to the dependence of the PDF uncertainty on the different flavors in the proton and on the most problematic range in partonic-$x$.
\begin{table}[!h]
\begin{center}
\begin{tabular}{|c|c|c|c|}
\hline
\multicolumn{4}{|c|}{normalized distributions} \\
\hline
cut on $p_\perp^W$ & cut on $|\eta_l|$   & CT10                      & NNPDF3.0 \\
\hline\hline
inclusive         & $|\eta_l|<2.5$      & $80.400 + 0.032 - 0.027$ & $80.398 \pm 0.014$  \\
\hline
$p_\perp^W <20$ GeV & $|\eta_l|<2.5$     & $80.396 + 0.027 - 0.020$  & $80.394 \pm 0.012$ \\
\hline
$p_\perp^W <15$ GeV & $|\eta_l|<2.5$     & $80.396 + 0.017 - 0.018$ & $80.395 \pm 0.009$  \\
\hline
$p_\perp^W <10$ GeV & $|\eta_l|<2.5$     & $80.392 + 0.015 - 0.012$  & $80.394 \pm 0.007$ \\
\hline\hline
$p_\perp^W<15$ GeV & $|\eta_l|<1.0$      & $80.400 + 0.032 - 0.021$ & $80.406 \pm 0.017$  \\
\hline
$p_\perp^W <15$ GeV & $|\eta_l|<2.5$     & $80.396 + 0.017 - 0.018$ & $80.395 \pm 0.009$  \\
\hline
$p_\perp^W<15$ GeV & $|\eta_l|<4.9$      & $80.400 + 0.009 - 0.004$ & $80.401 \pm 0.003$ \\
\hline
$p_\perp^W<15$ GeV & $1.0 < |\eta_l|<2.5$ & $80.392 + 0.025 - 0.018$ & $80.388 \pm 0.012$  \\
\hline
\end{tabular}
\caption{
LHC 8 TeV, $W^+$ production. Impact of different acceptance cuts.
The two cuts $p_\perp^l>25$ GeV and $\rlap{\slash}{\! E_T} \geq 25$ GeV are always applied.
In the first four rows we vary the cut on $p_\perp^W$, for fixed $|\eta_l|$ interval. In the second four rows we vary the pseudorapidity acceptance, with $p_\perp^W<15$ GeV.
\label{tab:cuts}
}
\end{center}
\end{table}
\begin{figure}[!h]
\begin{center}
\includegraphics[width=80mm,angle=0]{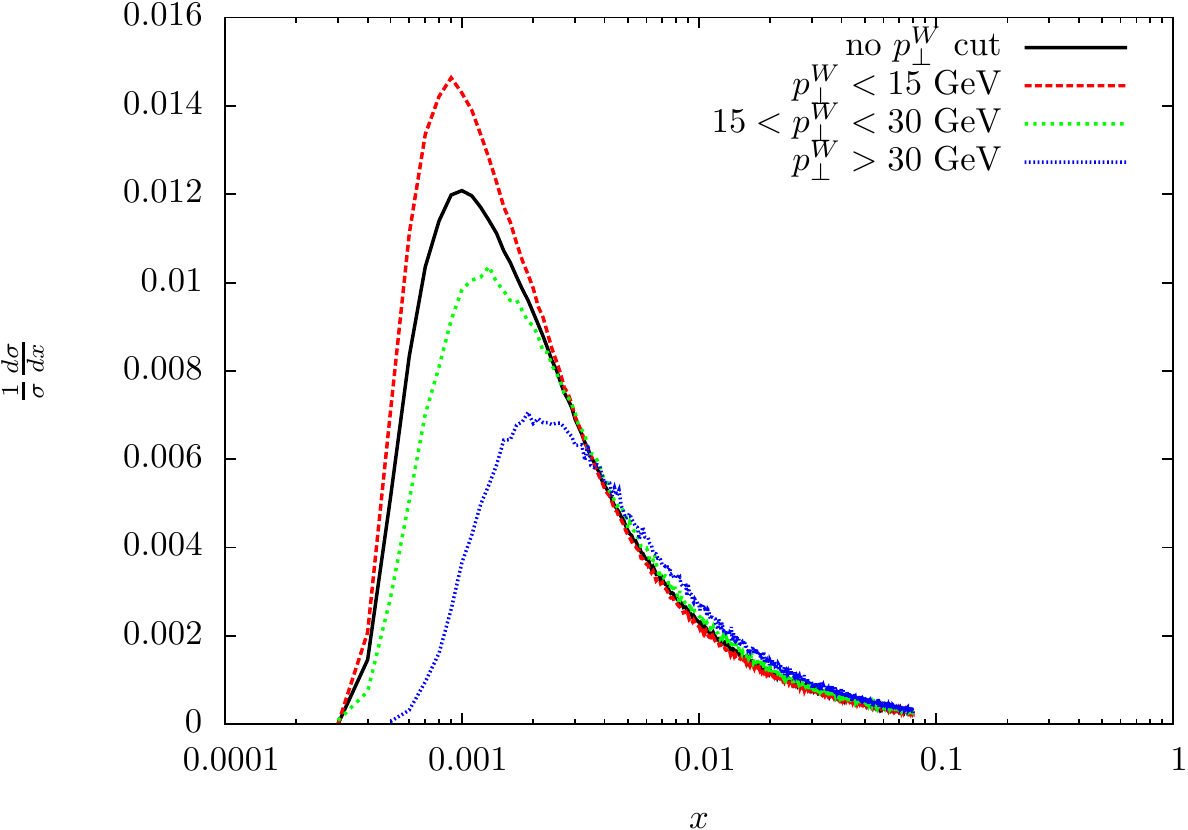}
\includegraphics[width=80mm,angle=0]{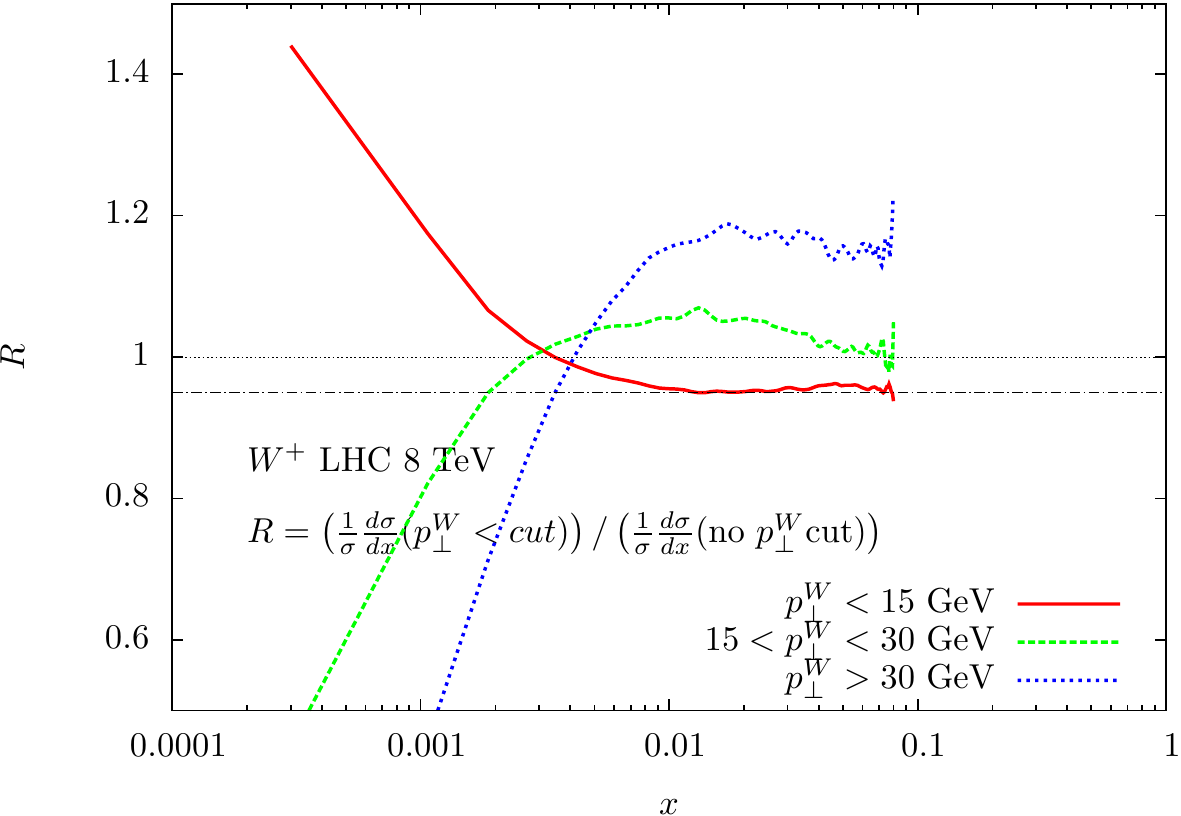}
\caption{
Shape of the normalized differential distribution $d\sigma/dx$ for different $p_\perp^W$ cuts (left plot).
Ratio of the previous shapes with different $p_\perp^W$ cuts with respect to the inclusive (no $p_\perp^W$ cut) distribution (right plot).
\label{fig:pdfunccuts1}
}
\end{center}
\end{figure}
\begin{figure}[!h]
\begin{center}
\includegraphics[width=80mm,angle=0]{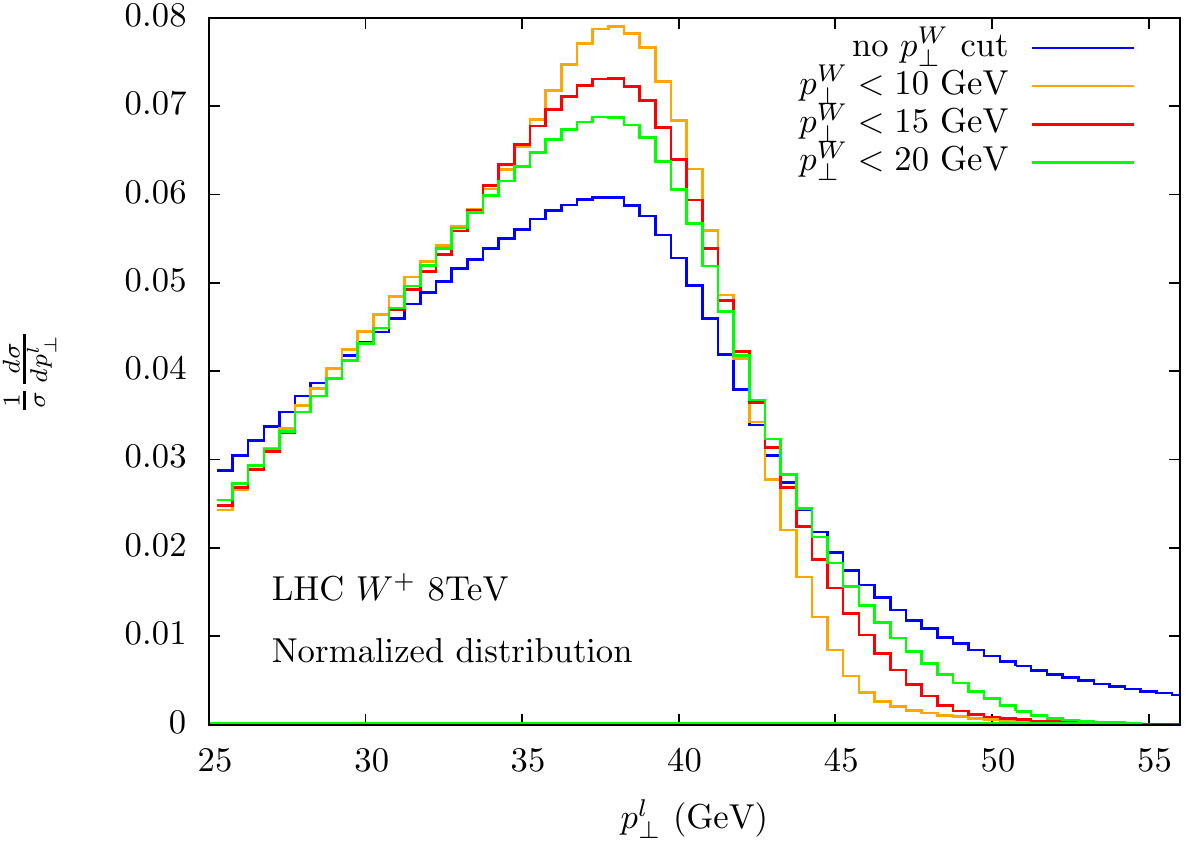}
\includegraphics[width=80mm,angle=0]{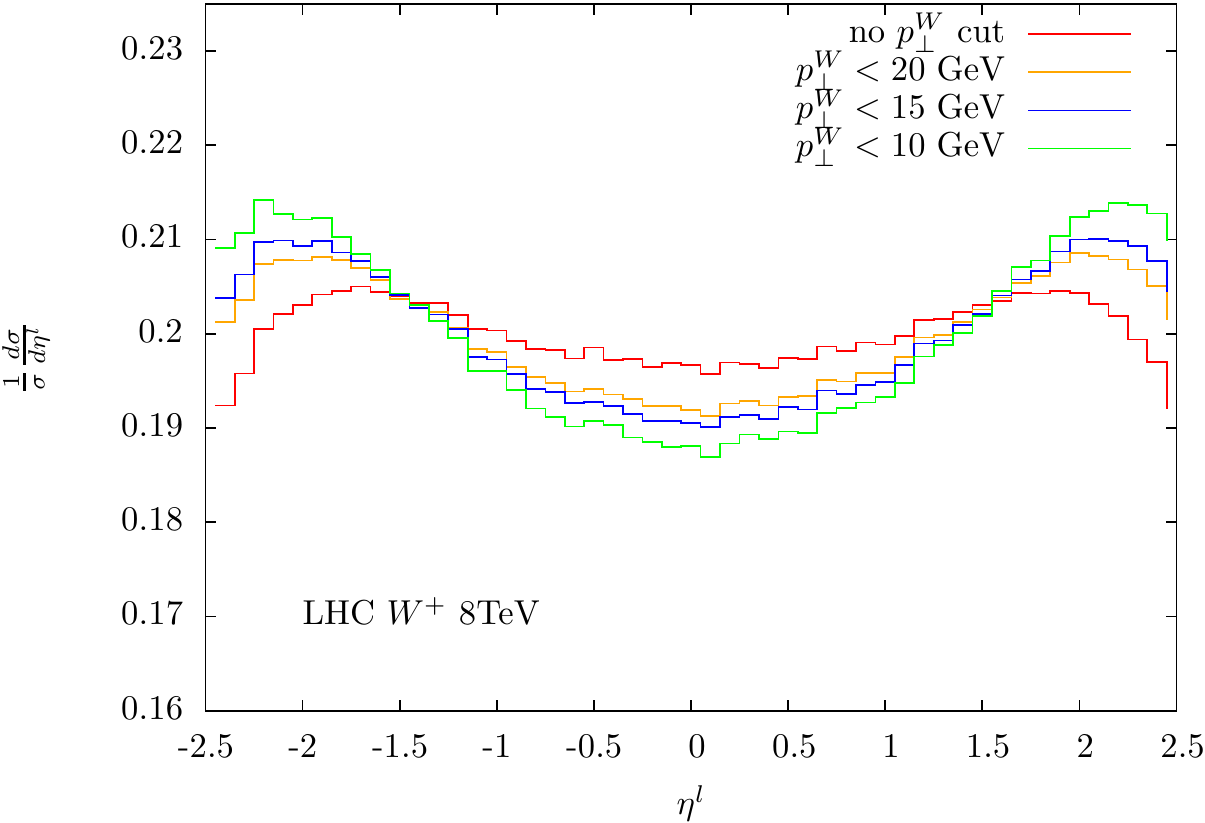}
\caption{
Shape of the lepton transverse-momentum (left panel) and of the lepton pseudorapidity (right panel) distributions, in presence of different 
additional cuts on the lepton-pair transverse momentum $p_\perp^W$.
\label{fig:ptlslope}
}
\end{center}
\end{figure}

\begin{figure}[!h]
\begin{center}
\includegraphics[width=80mm,angle=0]{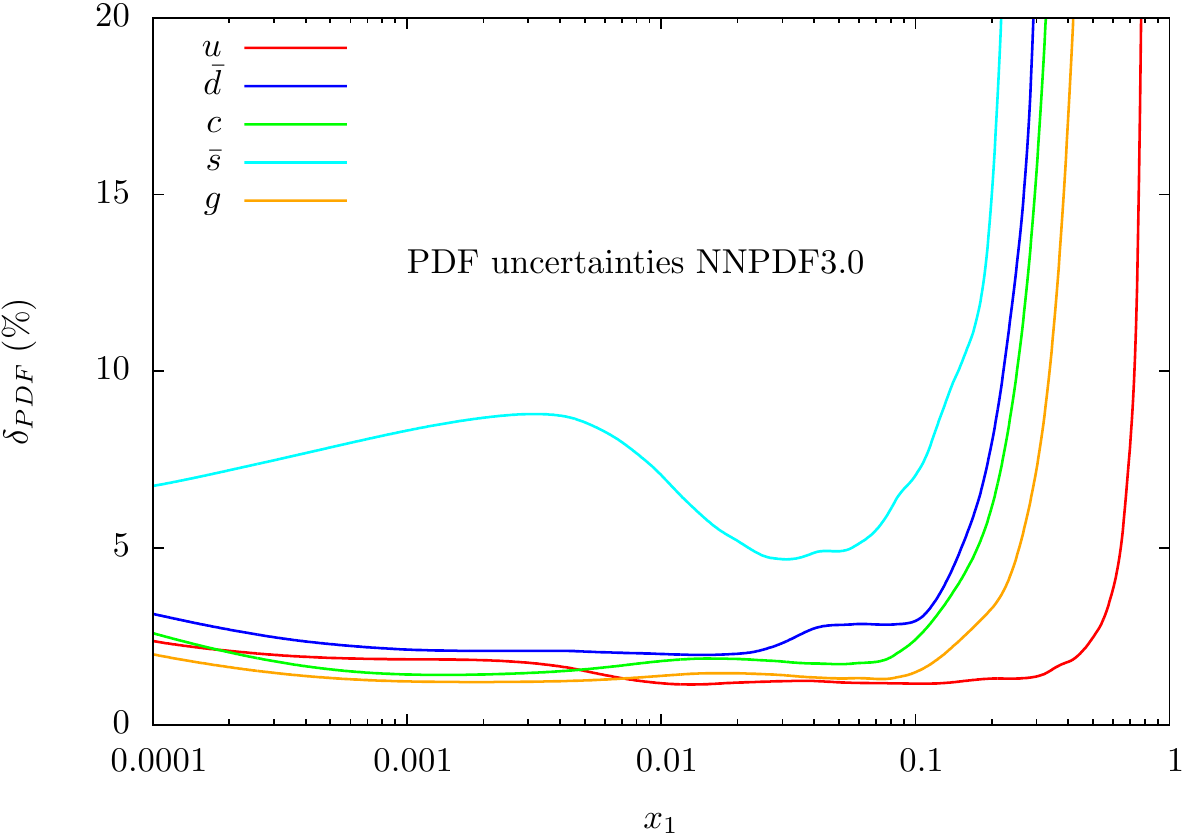}
\includegraphics[width=80mm,angle=0]{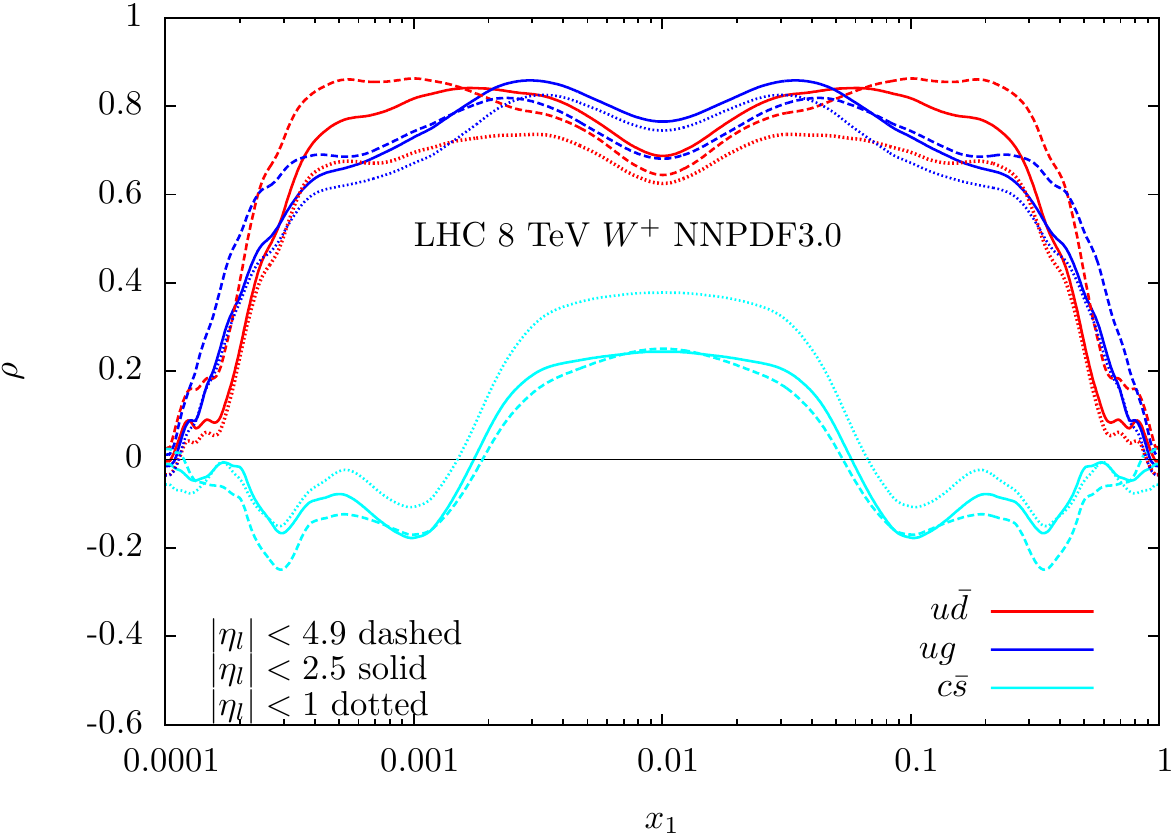}
\caption{
Percentage uncertainty of the individual parton densities $f(x,\mw^2)$ of {\tt NNPDF3.0}
(left plot).
Correlation of different parton-parton luminosities with the charged-lepton $p_\perp^l$ distribution at $p_\perp^l=40.5$ GeV, computed with different acceptance cuts on $|\eta_l|$ and with $p_\perp^W<15$ GeV.
\label{fig:pdfunccuts2}
}
\end{center}
\end{figure}
\begin{figure}[!h]
\begin{center}
\includegraphics[width=80mm,angle=0]{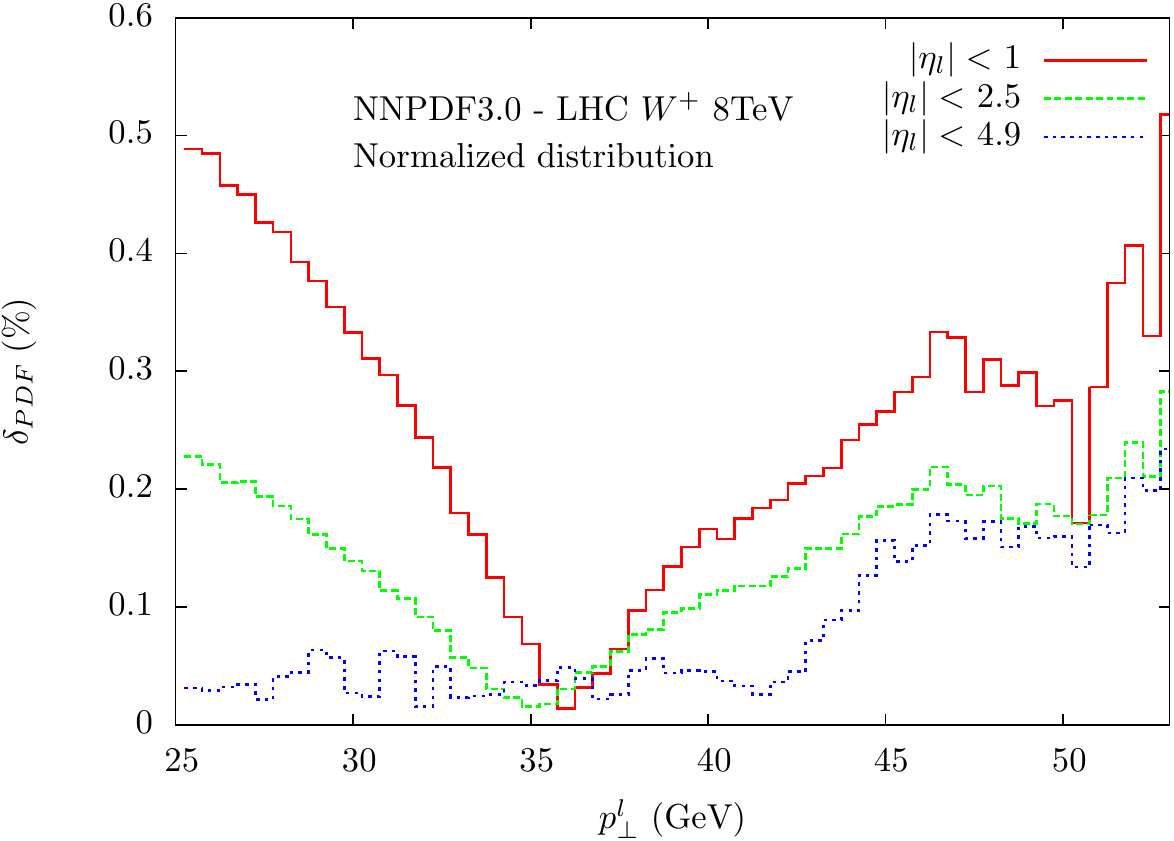}
\includegraphics[width=80mm,angle=0]{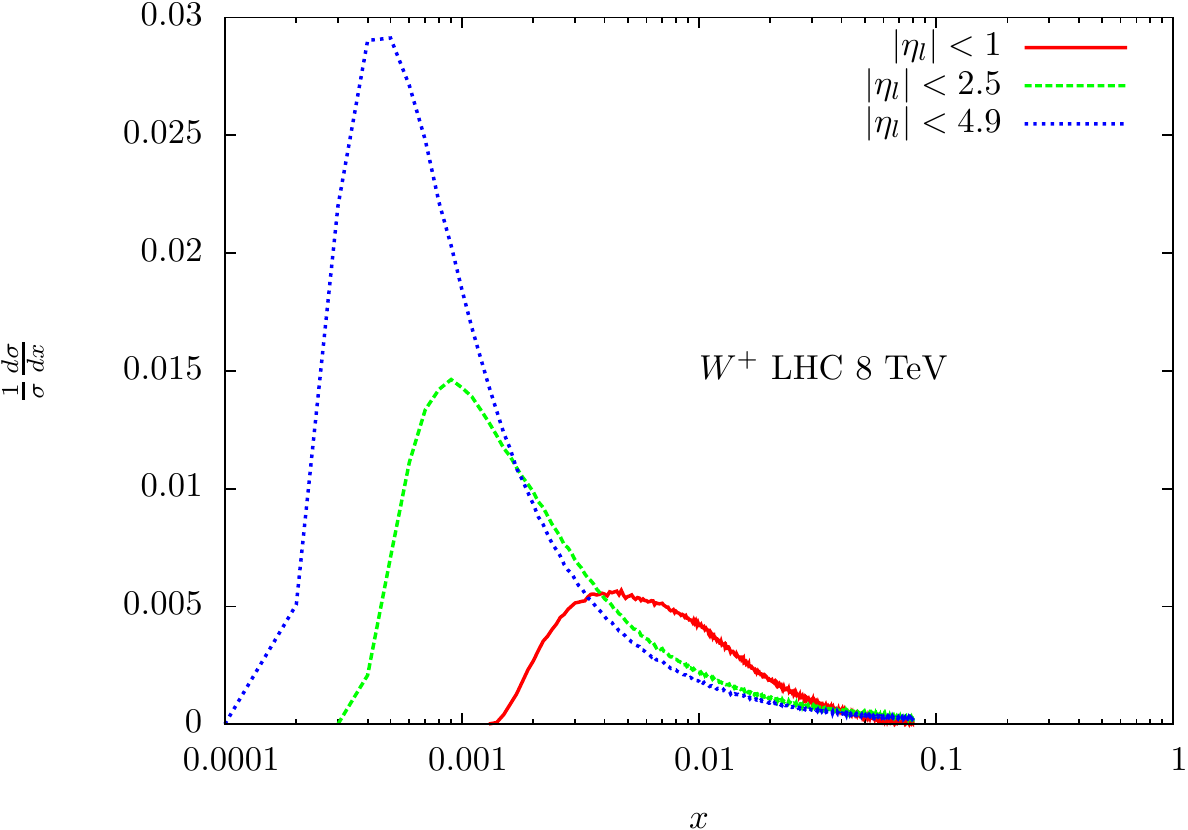}
\caption{
Percentage PDF uncertainty of the charged-lepton $p_\perp^l$ distribution (left plot) and 
shape of the differential distribution $d\sigma/dx$ (right plot),
computed with different  acceptance cuts on $|\eta_l|$ and with $p_\perp^W<15$ GeV.
\label{fig:pdfunccuts3}
}
\end{center}
\end{figure}
We observe that the region at large $p_\perp^W$ yields an important contribution to the PDF uncertainty, which can be reduced by a suitable cut on this variable. A tight cut like $p_\perp^W<10$ GeV could bring the uncertainty below the 10 MeV level.
The experimental problem to accurately select the events that pass the cut can be a limiting factor for the improvement in this direction.

The impact of the cut on the lepton-pair transverse momentum can be explained by studying the change of the relative contribution of the medium- $vs$ the large-$x$ PDF region, where $x$ is the fraction of momentum of the parent hadron carried by the incoming parton.
In Figure \ref{fig:pdfunccuts1} (left plot) we show the normalized $d\sigma/dx$ distributions, where $x$ is the fraction of longitudinal momentum carried by the partons of one given hadron in the 
scattering\footnote{The choice of the hadron is not relevant, because the 
contribution of the partonic subprocesses is symmetric for exchange of hadrons 1 and 2}; they are
computed with different $p_\perp^W$ cuts,
and express the relative contribution of a given partonic $x$ to the cross section.
In Figure \ref{fig:pdfunccuts1} (right plot) we show the ratio of the previous distributions, computed with different $p_\perp^W$ cuts,
with respect to the inclusive (no $p_\perp^W$ cut) normalized distribution. These ratios express the relative change of the weight of the various $x$ intervals, in presence of a cut.
We thus recognize that the $p_\perp^W<15$ GeV cut enhances the $x<0.004$ region and suppresses the contribution at $x>0.004$.
Since the  PDF uncertainty of all the densities rapidly increases for $x>0.1$ (cfr. Figure \ref{fig:pdfunccuts2}, left plot), 
the effect of the $p_\perp^W$ cut is a reduction of the global PDF uncertainty affecting the $\mw$ determination.
A second effect of the cut is a change of the basic shape of the distribution, which becomes steeper and closer the LO one, above the jacobian peak, as it is shown in Figure \ref{fig:ptlslope}: 
this modification increases the sensitivity of the fitting procedure, which becomes more stable, because large shifts are more penalized with respect to the case of a broader distribution.
In right panel of Figure \ref{fig:ptlslope} we show the normalized lepton pseudorapidity distribution, computed for different values of the $p_\perp^W$ cut. We observe that with tighter cuts the distribution develops two peaks at forward and backward rapidities.
These regions are dominated by the contribution of at least one valence quark, whose PDF uncertainty is smaller than the one of the corresponding sea component.

We observe that, for fixed cut on $p_\perp^W$, the PDF uncertainty decreases from 17 (26) to 3 (6) MeV with {\tt NNPDF3.0} ({\tt CT10}),
as one enlarges the charged-lepton rapidity cut, from 1.0 to 4.9.
This reduction is consistent with the smaller PDF uncertainty of the lepton transverse momentum distribution with the cut $|\eta_l|<4.9$ shown in Figure \ref{fig:pdfunccuts3} (left plot).
In this case the problematic point is the possibility of an accurate measurement of the lepton properties in the large rapidity regions of the detector.

The impact of the cut on the charged-lepton pseudorapidity can be explained 
first of all by recalling that a lepton transverse momentum distribution fully integrated over the lepton-pair rapidity (without acceptance cuts)
would depend on the PDFs only via a single numerical factor, which drops out when we study the normalized distributions.
This ideal limit can be reached, in a realistic setup,
by enlarging the charged-lepton pseudorapidity acceptance.
More in detail, with different maximal values of $\eta_l$,
we observe a corresponding change of the shape of the $d\sigma/dx$ distribution,
shown in Figure \ref{fig:pdfunccuts3} (right plot): the bulk of the distribution is peaked around $5\cdot 10^{-3},\, 1\cdot 10^{-3},\,5\cdot 10^{-4}$ respectively for $|\eta_l|<1,\,2.5,\,4.9$.

First of all we observe in Table \ref{tab:cuts}
that the two PDF uncertainties on $\mw$ extracted imposing the cuts
$|\eta_l|<1.0$ or $1.0< |\eta_l|<2.5$ are separately larger than the one obtained with $|\eta_l|<2.5$. Indeed, the sensitivity to partonic $x$ obtained by varying the cut on $\eta_l$ makes evident the presence of the momentum sum rules, which have to be fulfilled by all the replicas: 
the more inclusive setup is thus more stable with respect to a PDF replica variation than the more exclusive cases.
This uncertainty reduction is even more pronounced with $|\eta_l|<4.9$.

Second, in the region $5\cdot 10^{-3}\le x \le 1\cdot 10^{-2}$ the strange density has its maximal uncertainty, which is more than 3 times larger than the one of all the other parton densities, as shown in \ref{fig:pdfunccuts2} (left plot); in this same region the parton-parton luminosity $c\bar s$
has a weak positive correlation with respect to the PDFs,
defined in Equation \ref{eq:correlation}, 
with the charged-lepton transverse momentum distribution at $p_\perp^l=40.5$ GeV, so that its contribution to the cross section and to the total PDF uncertainty sums together with the ones of the other channels.
In the interval $3\cdot 10^{-4}\le x \le 7\cdot 10^{-4}$ the strange density still has a PDF uncertainty 2.5 times larger than the others, 
but in this region the $c\bar s$ luminosity has a negative correlation with the distribution.
In this case there are non trivial compensations between the contributions of the various partonic subprocesses, yielding a more stable result with respect to the PDF replica choice. 
This behavior of the parton densities,
in the case of this quite inclusive observable, with the acceptance cut $|\eta_l|<4.9$,
reflects the enforcement in the global fit of the sum rules that have to be satisfied by the PDFs.

\subsection{Comparison with previous studies}
\label{sec:previous}
The PDF uncertainty affecting the $\mw$ determination from the study of the lepton transverse momentum distribution
has been estimated in \cite{Aaltonen:2013vwa} and in \cite{D0:2013jba} to be respectively 12 and 11 MeV.
This evaluation was based on the simulation code {\tt ResBos} and on the use of the PDF sets {\tt cteq6.6} \cite{Nadolsky:2008zw} and {\tt MSTW2008} \cite{Martin:2009iq}.
We repeated the estimate of the uncertainty in the Tevatron setup also with
 {\tt cteq6.6},
using {\tt POWHEG}+{\tt PYTHIA} and templates computed with {\tt NNPDF2.3} replica 0; we obtain $\mw = 80.396 + 0.015 - 0.016$, i.e. a slightly larger PDF uncertainty compared to the previous estimate.

At variance with the transverse mass case \cite{Bozzi:2011ww}, 
the study presented in this note, with events treated at generator level,
is moderately sensitive to detector effects
and should thus represent a realistic estimate of the overall size of the PDF 
uncertainty and of the relative behavior of the different PDF sets.

\subsection{The role of the lepton-pair transverse momentum distribution in NC-DY}
The description of the lepton transverse momentum distribution
depends on the treatment of initial-state QCD radiation, to obtain a correct lepton-pair transverse-momentum distribution and in turn the correct contribution to the lepton transverse momentum.
At low lepton-pair transverse momenta there are non-vanishing non-perturbative effects, which can be accounted for by means of {\it ad hoc} models, upon which the final result of $\mw$ depends.
The uncertainties of the PDFs and of the modeling of an intrinsic component $k_\perp$
of the transverse momentum of the partons inside the proton are entangled in the lepton-pair transverse momentum distribution, because of the different contribution of the various flavors to the transverse momentum spectrum; in other words, it is not possible, in principle, to derive a universal, flavor independent, model of the intrinsic $k_\perp$.
This statement has been investigated in the past (see e.g. \cite{Konychev:2005iy})
where the Tevatron data were described by a universal flavor-independent non-perturbative functions.

A reduction of this dependence can be obtained by considering new observables, defined as ratios of the CC-DY observables with respect to their analogous ones in the case of NC-DY \cite{Giele:1998uh}.
The similarities in the initial-state QCD radiation patterns
determine a correlation between the CC-DY and the NC-DY quantities, 
which in turn yields a reduction of the error that affects the ratio.
One should however keep in mind that it is not possible to expect a full correlation between the
CC-DY and the NC-DY observables, because of the different flavor structure of the subprocesses in the two cases and because of the different phase-spaces available.

Given the entanglement between PDF and intrinsic $k_\perp$ uncertainties,
the estimate of the PDF uncertainty alone presented in this paper, for the CC-DY case,
could be a slight overestimate of its contribution to the total non-perturbative uncertainty.
On the other hand, the estimate of the PDF uncertainty for the ratio of CC-DY with respect to NC-DY observables should offer a more reliable result,
thanks to the weaker model dependence.
A detailed study of these ratios and of the associated theoretical uncertainties will be presented elsewhere.

\section{Conclusions}
\label{sec:conclusions}
We presented a quantitative assessment of the PDF uncertainty affecting the extraction of the $W$ boson mass from the study of the charged-lepton transverse-momentum distribution in the charged-current Drell-Yan process,
at different hadron colliders, for different collider energies.
The study, conducted at generator level, is based on the Monte Carlo code
{\tt POWHEG} interfaced with the {\tt PYTHIA} QCD parton shower
and uses the {\tt NNPDF2.3} PDF set (replica 0) to prepare the templates
used in the fitting procedure.
The results are summarized in Figure \ref{fig:summaryunc} and in Table \ref{tab:spread}.
The study provides information about the relative distance between the {\tt NNPDF2.3} and the other sets considered ({\tt CT10, MSTW2008CPdeut});
this distance is expressed by the difference between the best predictions of the various sets and ranges between 8 and 15 MeV, depending on the collider, on the energy and on the final state considered; these results rely on the application of a cut on the lepton-pair transverse momentum, $p_\perp^W<15$ GeV.
The study provides an estimate of the PDF uncertainty according to the prescriptions of each PDF group: the individual values range between 6 and 18 MeV, again depending on the considered setup and always in presence of the cut on the lepton-pair transverse momentum.
The combination of the two previous uncertainties, according to the PDF4LHC recipe, leads to a global PDF uncertainty that ranges between 18 and 24 MeV.
The analysis of more modern sets, like {\tt NNPDF3.0} and {\tt MMHT2014},
does not change this overall picture, but makes evident some differences in the description of $W^+$ with respect to $W^-$ production.

We remark that the differences between the PDF sets considered here
are large compared to an accuracy goal of 10 MeV in the $\mw$ measurement.
On the other hand, the fact that the individual sets predict uncertainties in the 10 MeV ballpark leaves hope that an improvement of the global PDF analysis will remove this bottleneck towards a precise $\mw$ measurement.

The variation of the acceptance cut on the lepton pseudo-rapidity
offers the possibility to scrutinize the dependence of the uncertainty on the flavor content of the proton and on the values of partonic-$x$.
The preliminary results are not trivial, because of the correlations among the densities enforced by the PDF sum rules. Increasing the value of the cut on  $|\eta_l|$ reduces the PDF uncertainty on $\mw$.

\section{Acknowledgements}
AV would like to thank 
Stefano Forte, Joey Huston and Robert Thorne for comments about the PDF sets, 
Maarten Boonekamp and Luca Perrozzi for useful information about the experimental analyses,  
Michelangelo Mangano for an interesting conversation and 
Paolo Nason for  an important clarification about {\tt POWHEG}.
AV is supported in part by an Italian PRIN2010 grant, by a European Investment Bank EIBURS grant, and by the European Commission through the HiggsTools Initial Training Network PITN-GA-2012-316704.

\end{document}